\documentclass{aa}  

\usepackage{graphicx}
\usepackage{gensymb}
\usepackage{verbatim}
\usepackage{siunitx}
\usepackage{amsmath}
\usepackage{caption, subcaption}
\usepackage[colorlinks,allcolors=blue]{hyperref}
\usepackage{txfonts}
\usepackage{amssymb}
\usepackage{chemformula}
\usepackage{multirow}
\usepackage{subcaption}
\graphicspath{{figures/}}
\usepackage{placeins}

\def \HII {H{\,\textsc{ii}} }

\def \cCI {[C\,{\sc{i}}] }
\def \CI {[C\,{\sc{i}}] } 
\def \cCII {[C\,{\sc{ii}}] }
\def \13cCII {[$^{13}$C\,{\sc{ii}}] } 
\def \CII {[C\,{\sc{ii}}] }

\def \kms {km s$^{-1}$}

\newcommand{\deceased}{\textsuperscript{\textdagger}}

\newcommand{\mybold}[1]{\textbf{#1}}
\renewcommand{\mybold}[1]{#1}

\newcommand{\blfootnote}[1]{%
  \begingroup
  \renewcommand{\thefootnote}{}%
  \footnotetext{#1}%
  \endgroup
}

\begin{document}

   \title{Multi-phase Carbon Structure and Dynamics in NGC 6334: Insights from CO, [C I], and [C II]}   

   \author{S. Neupane \inst{1}
   \and F. Wyrowski \inst{1}
   \and N. Schneider \inst{2} 
   \and A.G.G.M. Tielens \inst{3,4}
   \and R. Simon \inst{2}
   \and J.\,S.\, Urquhart \inst{5}
   \and R. G\"usten \inst{1}
   \and M-Y. Lee \inst{6}
    \and D. Russeil \inst{7}
    \and G. Garay \inst{8,9}
    \and R. Higgins \inst{2} 
    \and K. M. Menten \inst{1}\deceased
   }
   \institute{Max-Planck-Institut f\"{u}r Radioastronomie, Auf dem H\"{u}gel 69, 53121 Bonn, Germany 
              \email{sneupane@mpifr-bonn.mpg.de}
              \and 
		Institut f\"{u}r Astrophysik, Universit\"at zu Köln, Z\"ulpicher Str. 77, 50937 K\"oln, Germany
        \and 
        University of Maryland, Department of Astronomy, College Park, MD 20742-2421, USA
        \and 
        Leiden Observatory, PO Box 9513, 2300 RA Leiden, The Netherlands
        \and 
        Centre for Astrophysics and Planetary Science, University of Kent, Canterbury, CT2 7NH, UK 
        \and 
        Korea Astronomy and Space Science Institute, 776 Daedeokdae-ro, Daejeon 34055, Republic of Korea
        \and 
        Aix Marseille Univ., CNRS, LAM, Laboratoire d’Astrophysique de Marseille, 13013 Marseille, France
        \and 
        Departamento de Astronomía, Universidad de Chile, Casilla 36-D, Santiago, Chile 
        \and
        Chinese Academy of Sciences South America Center for Astronomy, National Astronomical Observatories, CAS, Beijing 100101, PR China
        }

   \date{Received May 2026; Accepted Nov 2026}
 
  \abstract
   { 
We present high-angular and velocity resolution observations of the emission lines of \cCII at 158\,$\mu$m, \cCI $^3P_1$–$^3P_0$ at 609\,$\mu$m, and the $^{12}$CO and $^{13}$CO $J=3\to2$ transitions toward the central region of NGC~6334, providing a detailed multi-phase view of carbon in this massive star-forming complex. By combining morphology and spatial scales information, kinematics, line profiles, and line ratio analyses, we quantify the distribution and physical state of ionized and atomic carbon, and carbon in its molecular form in CO. \CII emission is the most extended, permeates the mapped region and is dynamically complex, while $^{13}$CO is compact and quiescent, and $^{12}$CO and \CI occupy intermediate regimes connecting dense regions to transitional diffuse layers. We interpret these emission distributions such that the brightest \CII emission traces photodissociation regions (PDRs) close to the \HII region bubbles located along the central filament, while the extended \CII emission arises from the more diffuse warm gas, indicating either cloud formation or dispersal processes. Remarkably, the \CI is more extended than the $^{13}$CO, but less than \CII and may probe the gas velocity structure resulting from colliding flows or cloud-cloud collisions. Line ratios involving \CII, such as \cCII/\cCI or \cCII/CO are systematically high, revealing widespread diffuse emission and localized UV-illuminated surfaces. Line ratio analysis and K-means clustering of \cCI/$^{13}$CO versus dust continuum (cold 870 $\mu$m and warm 70 $\mu$m) emission reveal stratified chemical structure that UV photons penetrate deeply into the clumpy ridge ($A_V \sim 18$), suggesting a porous PDR. These results together depict that NGC~6334 is a clumpy, feedback-dominated PDR ridge with coexisting carbon phases, where \cCI and \cCII also trace dynamical processing through multiple velocity components linked to cloud–cloud collisions.}


\authorrunning{Neupane et al.}
\titlerunning{Multi-phase Carbon Structure and Dynamics in NGC 6334}

\maketitle

\blfootnote{\deceased Deceased [30 Dec 2024]. In memoriam and with gratitude.}
\nolinenumbers 

\section{Introduction}
\begin{figure*}[hbtp!]
    \centering
    \includegraphics[width=0.75\linewidth]{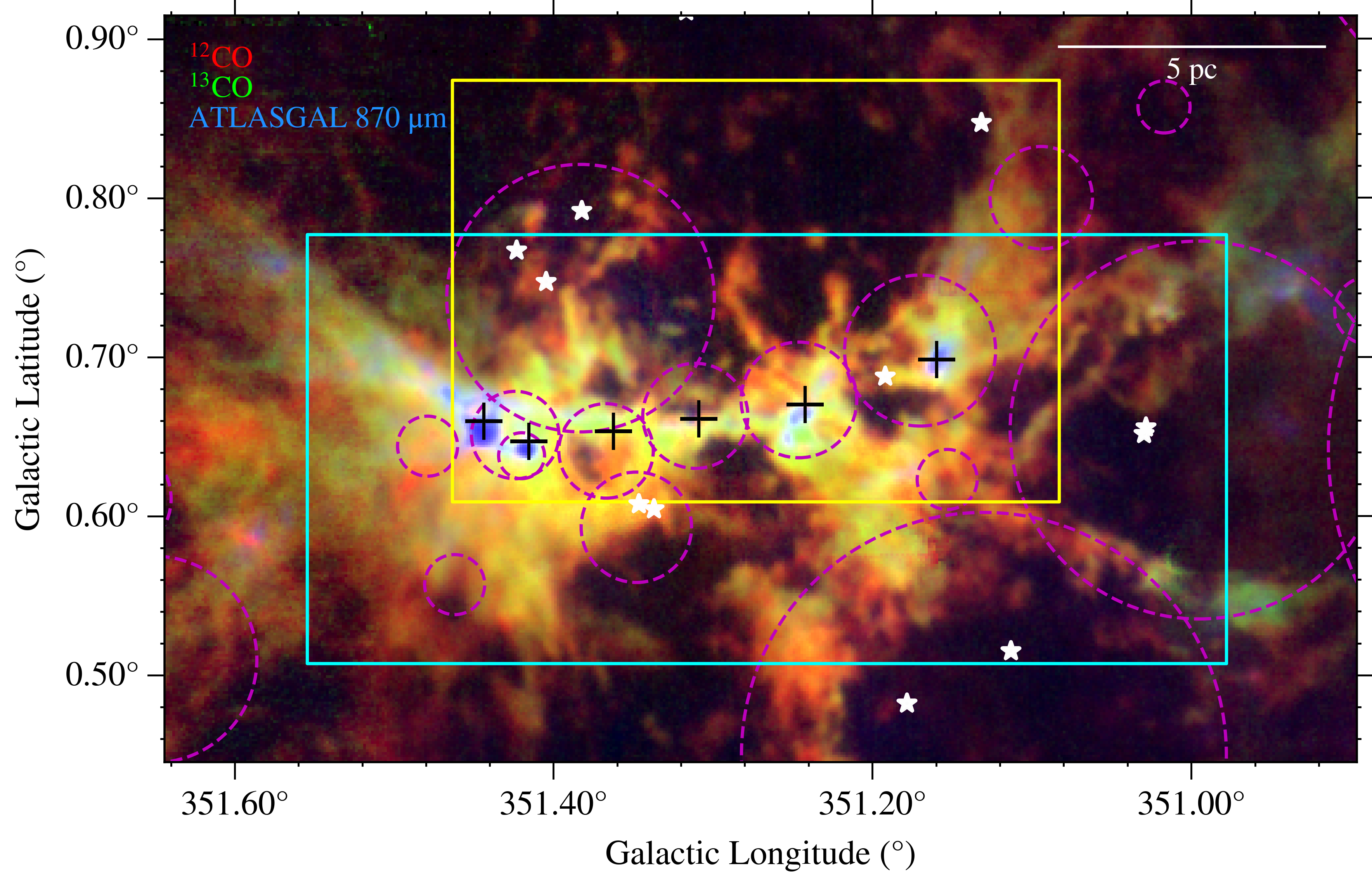}
    \caption{Three color image of the NGC 6334 central filament. Red: $^{12}$CO, Green: $^{13}$CO, Blue: ATLASGAL 870 $\mu$m emission. The \HII regions from the \cite{Anderson2014} are indicated by the dashed magenta circles. Six far-infrared (FIR) sources (I[N], I to V from left to right; for more details see \citealt{Kraemer+Jackson1999}) located in the central filament are indicated by the black plus signs. White star symbols indicate the OB--type stars from \cite{persitapia2008}. The cyan and yellow rectangular boxes outline regions covered in the \cCI and \cCII lines, respectively.} 
    \label{fig:mom0map1}
\end{figure*}

NGC~6334 is a star forming  complex  (of extent $\sim$80 pc), harbouring a central dense filament ($\sim$10 pc length) and a large number of OB-type stars, \HII regions and bubbles in the extended region (Fig. \ref{fig:mom0map1}) (e.g., \citealt{persitapia2008,Russeil2013,Tige2017,Louvet2019, Neupane2024}). This complex is located at a distance of 1.7 kpc (e.g., \citealt{Russeil2012}). The central dense filament contains sites of high-mass star formation in a sequence of evolutionary stages (e.g, \citealt{Tige2017}) and is proposed to undergo longitudinal collapse, possibly  triggered by a past high-mass star forming event (\citealt{Zernickel2013,Russeil2013,Neupane2024}). NGC~6334 has been reported to be undergoing large scale cloud-cloud collision by \cite{FukuiY2018CC}, supported by recent high-resolution observations of $^{12}$CO and $^{13}$CO $J=3-2$ that identified multiple velocity components and `bridge' features (\citealt{Neupane2024}). It has been proposed that the cloud-cloud interaction is happening over the past few million years triggering sequential star formation in the filamentary complex (\citealt{FukuiY2018CC}). 
Owing to the complex interplay of dense molecular cores, extended photon-dominated regions (PDRs), active feedback and dynamic interactions, NGC~6334 provides an ideal environment to investigate the origins of molecular, neutral, and ionized carbon emission.

The molecular, neutral and ionized carbon\footnote{C$^{0}$ or \CI and C$^{+}$ or \CII\ denote the species neutral and ionized carbon, and \cCII and \cCI are used for the spectroscopic notation of the fine-structure lines. 
Unless explicitly stated otherwise, we use \textsuperscript{12}CO and \textsuperscript{13}CO to refer to the $J\! =\! 3\!-\!2$ rotational transitions of $^{12}$CO and $^{13}$CO, respectively.} probe different layers of molecular clouds and PDRs. The \textsuperscript{12}CO and \textsuperscript{13}CO $J\!=\!3\!-\!2$ lines trace the colder, denser molecular interior, with \textsuperscript{13}CO providing a more optically thin probe of the cold and dense molecular gas (e.g., \citealt{Neupane2024}). Neutral carbon with its \cCI $^3\!P_{1}\!-\!^3\!P_{0}$ line  at $\nu = 492.160\,\mathrm{GHz}$ (609 $\mu$m) is thought to probe both molecular and neutral medium tracing the transitional layer (e.g., \citealt{Frerking1989, Clark2019}). Ionized carbon with the \cCII $^2\!P_{3/2}\!-\!^2\!P_{1/2}$ line at $\nu = 1900.536\,\mathrm{GHz}$ (158 $\mu$m) is dominant in the warm, UV-illuminated outer PDR layers of molecular clouds (e.g., \citealt{Tielens1985, Stacey1991, Madden2020}). Disentangling these multiple origins is challenging, as \cCII  emission traces a wide range of physical conditions in PDRs, from \HII region interiors to moderately dense cloud surfaces to extended diffuse atomic envelopes (e.g., \citealt{Pineda2013, Velusamy2015, Langer2021, Luisi2021, Schneider2023}).

High-resolution observations of CO, C$^{0}$, and C$^{+}$  are therefore essential to isolate spatial and kinematic components, constrain excitation conditions, and build a coherent understanding of carbon chemistry and thermal balance in the interstellar medium. Our aim in this paper is to investigate the spatial and velocity structure of these tracers in the central NGC~6334 filament (see Fig. \ref{fig:mom0map1}) using high-resolution data from the SOFIA and APEX telescopes, convolved to a common angular resolution of 20$^{''}$ ($\sim$0.16~pc at 1.7~kpc). The NGC 6334 central region was selected in the SOFIA FEEDBACK legacy project (\citealt{sofiafeedback}) because of its distinct environment as it harbors a large number of embedded and optical \HII\ regions in and around the central dense filament. NGC~6334 does not exhibit clearly defined large-scale expanding \cCII shells \citep{Faerber2025}, that are commonly detected in \HII\ region bubbles such as Veil bubble in Orion, RCW120, RCW49, NGC7538, and RCW36 \citep{Pabst2019, Tiwari2022, Beuther2022, Bonne2022}. 

This paper is organized as follows. Section~\ref{sec:observations} describes the observations and data reduction. Section~\ref{sec:results} presents the data, including representative spectra, emission morphology, and position–velocity diagrams. In Section~\ref{sec:analysis} we analyze the spatial structure using constrained diffusion decomposition, perform multi–Gaussian line fitting, examine line correlations, and derive key line ratios. Section~\ref{sec:discussion} discusses the physical implications of these findings, and Section~\ref{sec:summary} summarizes our conclusions.

\section{Observations}\label{sec:observations}
We mapped the extended region of NGC 6334 in the $^{12}$CO $J = 3 \rightarrow 2$ and $^{13}$CO $J = 3 \rightarrow 2$ molecular lines using the 7-pixel heterodyne array receiver, the Large APEX sub-Millimetre Array (LAsMA) (\citealt{ Gusten2008}), on the 12-meter single-dish Atacama Pathfinder EXperiment (APEX)(\citealt{Gusten2006}). The data reduction procedure is outlined in \cite{Neupane2024}, 
where we presented the molecular line emission properties and velocity structure of the region, covering an area of $2.5 \times 1.2$ square degrees. The spectral cubes have a resolution of 0.25 \kms, a typical rms of $\sim$0.40 K in one channel, and a full width at half maximum (FWHM) beam size of 20$''$.

The \CI observations at 492 GHz were conducted using the nFLASH460 instrument at the APEX telescope. We obtained On-The-Fly (OTF) maps covering the region of size $0.567 \times 0.372$ square degrees in the central NGC 6334 filament. 
While the native FWHM beam width of the data is $\sim$14$''$, we sampled the data to a spatial resolution of 20$''$ to match the resolution of the molecular CO data from the LAsMA observations. The spectra were smoothed to a resolution of 0.25 \kms, with a typical rms level of $\sim0.25$ K in one channel.

The ionized carbon \cCII  observations at 1.9 THz (or 158 $\mu$m) toward the central NGC 6334 filament were carried out using the upGREAT instrument (\citealt{upGREAT2016}) on SOFIA (Stratospheric Observatory for Infrared Astronomy) as part of the FEEDBACK project (\citealt{sofiafeedback}). Due to the shutdown of the observatory in 2022, only 36\% of the planned area could be observed, covering approximately an area of $0.37 \times 0.25$ square degrees. The observations were done as OTF maps with several coverages. For more observational details, see \citet{sofiafeedback}. The data reduction and smoothing were performed using the GILDAS{\footnote{https://www.iram.fr/IRAMFR/GILDAS/}} software package. The native FWHM beam width of the observations is $\sim$14$''$. We resampled the spectra to a common resolution of 0.25 \kms, achieving a typical rms of $\sim$0.67 K in antenna temperature, and smoothed the data to a spatial resolution of 20$''$ to match the resolution of the CO and \cCI line datasets mentioned above.

The mapped regions in \CI (using nFLASH/APEX) and \CII (using upGREAT/SOFIA) are indicated in Fig.~\ref{fig:mom0map1} by cyan and yellow rectangles, respectively. In this work, we study the central  NGC~6334 region of size $0.37 \times 0.25$ square degrees ($\sim$ 11.0~pc $\times$ 7.5~pc) where \CII observations are available. To obtain the corrected main beam temperature (T$_\mathrm{mb}$) from the antenna temperature (T$_\mathrm{A}$), we used beam efficiencies of 0.67, 0.60, and 0.72 for upGREAT/SOFIA, nFLASH, and LAsMA/APEX observations, respectively. 
We also complement these observations with continuum images, primarily using ATLASGAL 870 $\mu$m maps (\citealt{ATLASGAL2009}), Spitzer Space telescope (\citealt{Fazio2004}) emission maps at 8 $\mu$m (\citealt{Benjamin2003}) and at 70 $\mu$m (\citealt{Carey2009}). For a quantitative comparison of the fluxes, these emission maps were smoothed to a common resolution of $\sim$20$''$ and re-gridded to a pixel size of 6$''$. 

\begin{figure}[ht!]
    \centering
    \includegraphics[width=0.99\linewidth]{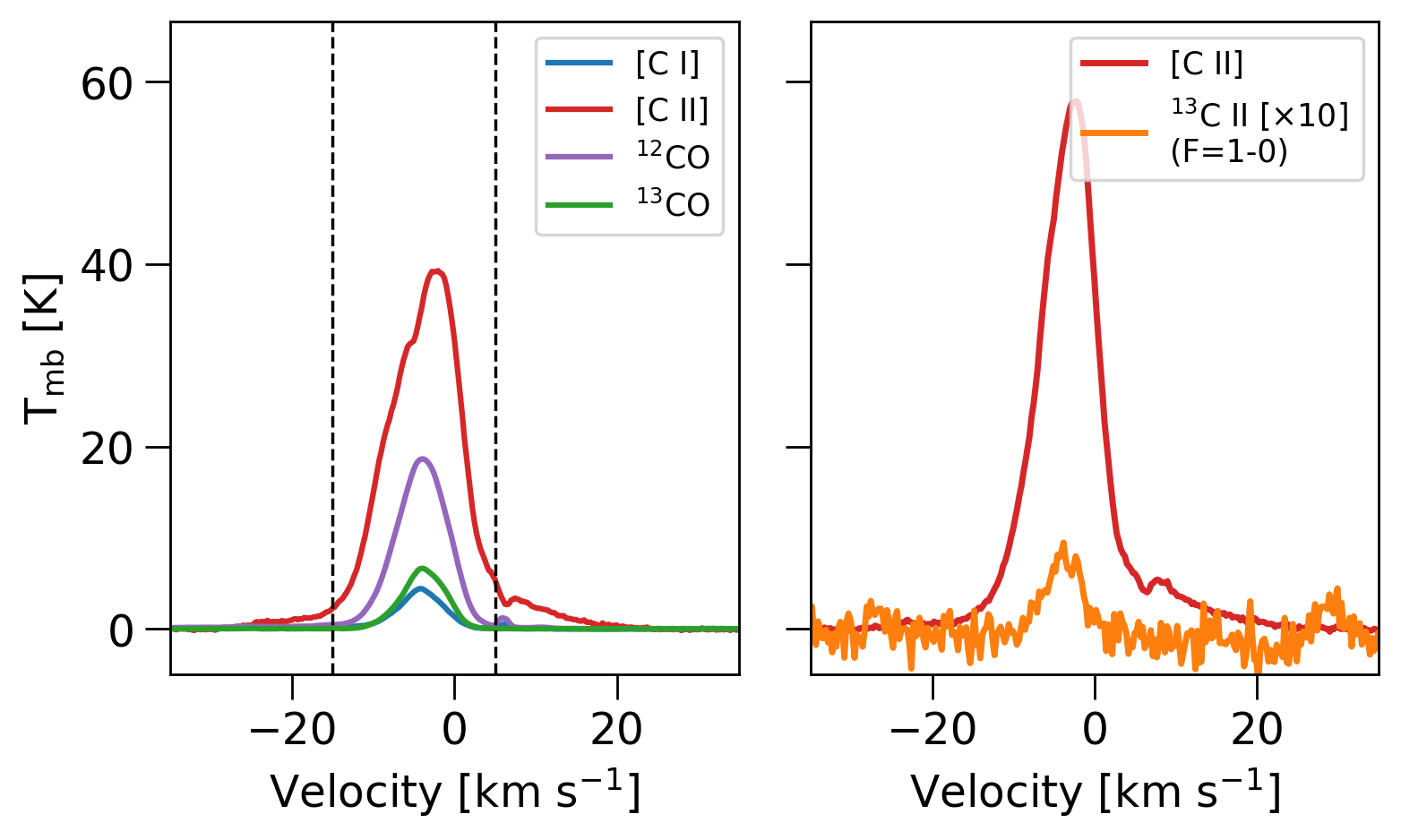}
    \caption{Spectra of four observed species. Left: Average spectra toward the common mapped region. The systemic velocity range $-$15 to +5 \kms\ is indicated by dashed black lines. Right: Average \cCII and \13cCII F=1-0 spectra toward the region indicated by the white dashed circle in Fig. \ref{momentmaps}. Peak values of \CII and \13cCII intensity are 54.77 K and 1.03 K, respectively, and the rms in the average spectrum is 0.12 K. }
    \label{fig:spectra}
\end{figure}

\section{Results}\label{sec:results}
\subsection{Spectra}\label{sec:spectra}

\begin{figure*}[htbp!]
    \includegraphics[width=0.99\linewidth]{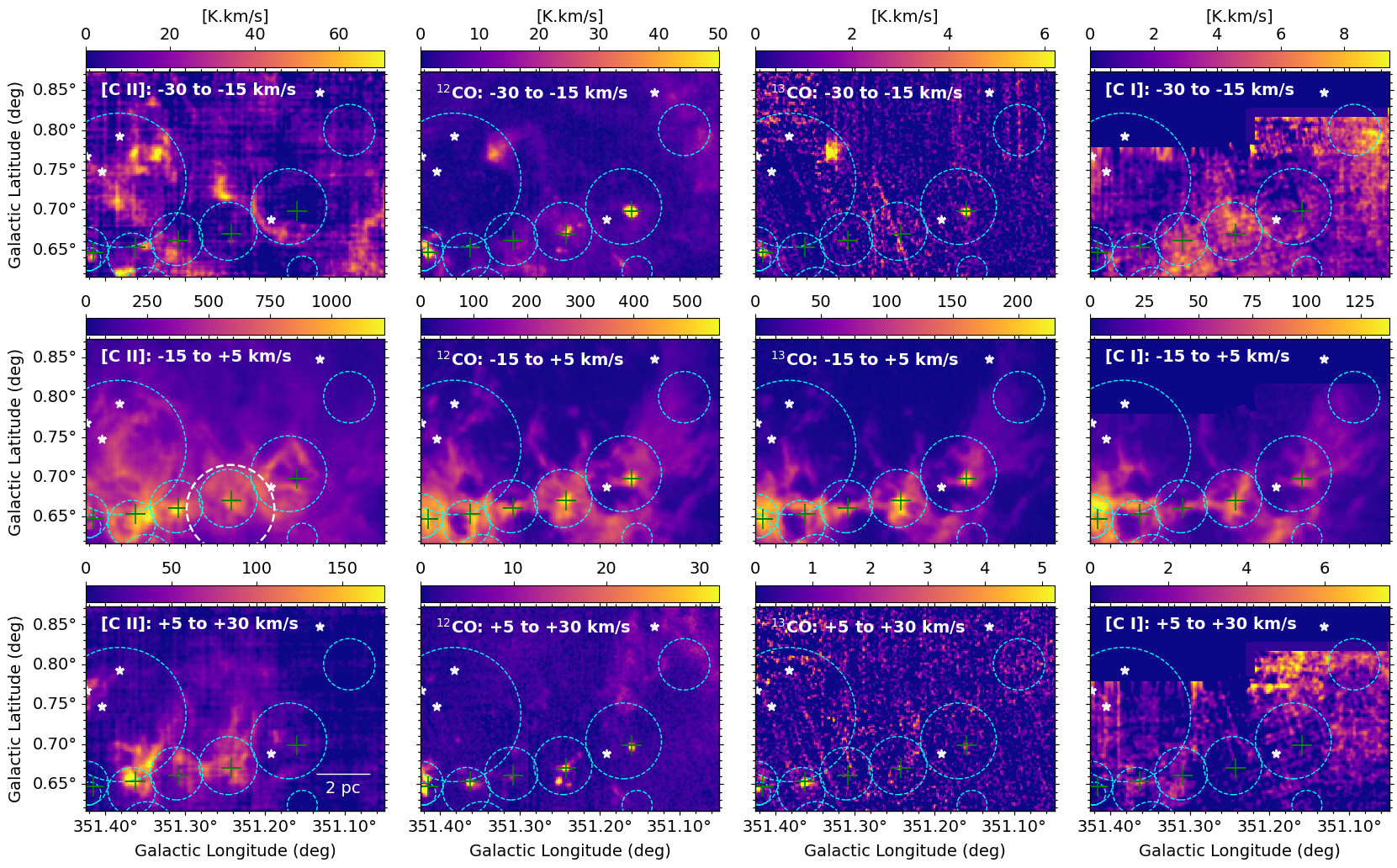}
    \caption{Moment-0 maps of the observed four lines. The maps display the main velocity ranges found in NGC6334. Contour levels are at 10, 30, 50, 70, 90 percent of the peak emission. In the \CII map, the region where the \13cCII F=1-0 line is detected in an averaged spectrum is outlined by a white dashed circle. Green plus markers on the maps indicate far-infrared sources I to V (\citealt{Kraemer+Jackson1999}) from left to right and white star symbols give the location of OB stars.  Magenta dotted circles show WISE \HII regions from \cite{Anderson2014}.} 
    \label{momentmaps}
\end{figure*}

Figure~\ref{fig:spectra} presents the average spectra of the four transitions in the common area (Fig.~\ref{fig:mom0map1}). The \cCII\ line is the brightest and exhibits pronounced high-velocity emission. The $^{12}$CO line is the second brightest, whereas $^{13}$CO and  the \cCI\ line display comparable brightness temperatures and similar spectral profiles. Emission near +7~\kms\ is detected in $^{12}$CO and \CI, consistent with a local velocity component, characterizing the foreground interstellar medium (historically identified by \cite{Heeschen1955, Riegel1972}, and mapped globally by \cite{Dame2001}). In the same velocity range, \CII\ shows an absorption feature that is likewise associated with the local emission.
The right panel in the figure presents spectra of the \13cCII hyperfine component  F=1-0. The region where this line is detected is indicated by a white dotted circle in Fig. \ref{momentmaps} (middle panel of \CII) and is not particularly associated with the location of the \HII regions. Outside of this area, \13cCII was not or only weakly detected (see below). Since \CII is widespread in velocity, the \13cCII F=2-1 component is blended with the \cCII emission and the \13cCII F=1-1 component is not detected as it has the lowest relative line strength. Relative line strengths of the lines are 0.250, 0.625 and 0.125, respectively for the F=1-0, 2-1 and 1-1 hyperfine lines (\citealt{Ossenkopf2013}). We estimate the optical depth of \CII using the \cCII and 
\13cCII F=1-0 peak intensity ratio and obtain a value of $\tau_\mathrm{CII}\sim$ 5, assuming $^{12}$C/$^{13}$C ratio of 67 (e.g, \citealt{Langer-Penzias1990,Milam2005,Kabanovic2022}). In this area the \cCII line is thus optically thick and line profiles show indications of self-absorption, pointing toward temperature gradients along the line-of sight \citep{Goldsmith2012,Kabanovic2022}. Interestingly, in the regions where \cCII is very bright, the \13cCII F=1-0 is not detected and we cannot derive the opacity of the \CII line (upper limit on $\tau_\mathrm{CII}<$15 for T$_\mathrm{peak}$ $>$55~K). However, T$_\mathrm{peak}$ in the central ridge is mostly similar or higher than the value of the averaged spectrum, so that \CII likely stays bright because of the high kinetic temperature. The line possibly remains optically thick in the central dense ridge and optically thin in the outskirts. This assumption is particularly valid if significant emission in \CII is coupled with dense gas tracers in the inside region producing optically thick emissions. 
From the observed average main beam brightness temperature of $\sim$55 K (Fig.~\ref{fig:spectra}), we obtain a Rayleigh-Jeans corrected temperature $T_{RJ}$ of 93 K using, 
\begin{equation}  \label{eq:RJ} 
T_{RJ} = \frac{h \nu}{k} \left[ \ln \left(1 + \frac{h \nu}{k T_{mb,C_{II}}}\right) \right] ^{-1},
\end{equation}
where, $h$ and $k$ denote the Planck and the Boltzman constants, respectively, and $\nu$ is the frequency. This is the excitation temperature for the \cCII\ line for regions of densities well above the critical density, below which it is the lower limit to the kinetic temperature (T$_{kin}$). We assume here that the densities are above the critical density of \CII for collisions with atomic and molecular hydrogen (3$\times$10$^3$ cm$^{-3}$ and 5-7$\times$10$^3$ cm$^{-3}$ at T$_{ex} = 100$ K, respectively (\citealt{Goldsmith2012}). \CII emission however can also arise from electron collision, particularly relevant for regions inside the \HII regions. Theoretical models suggest that the electron collision contribution in \CII emission is largely defined by the electron density, the size of \HII regions relative to the \CII emitting PDR layer and the stellar continuum \citep{Abel2006}. We briefly come back to the electron collision contribution in Section \ref{subsec:emissionmophology}.

In various parts of the analysis section presented below, we separate regions inside and outside of the 55\,K T$_\mathrm{mb}$ contour   (Figure~\ref{fig:ciitmbpeak}) to investigate properties of \CII bright vs less bright regions. 

\begin{figure*}[htbp!]
    \sidecaption
    \includegraphics[width=12cm]{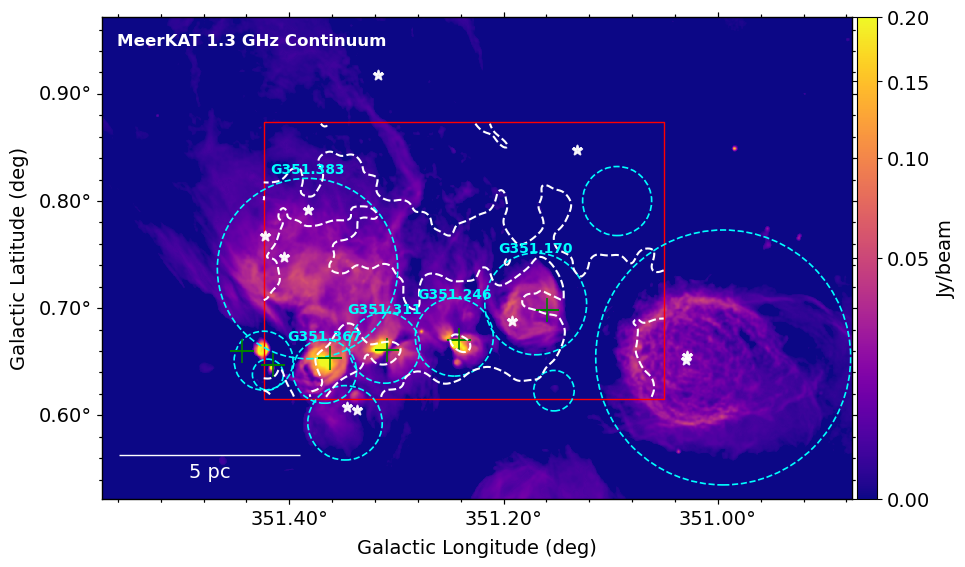}
    \caption{MeerKAT 1.3 GHz radio continuum  map overlayed with \cCII contours (at levels 200, 400 and 800 K \kms) in white dotted lines. The rectangle plotted in red color indicates our mapped region. Also shown are the \HII regions, OB stars and FIR sources as in Fig.~\ref{momentmaps}.}
    \label{ccontinuum}
\end{figure*}

\begin{figure*}[htbp!]
    \centering
    \includegraphics[width=0.99\linewidth]{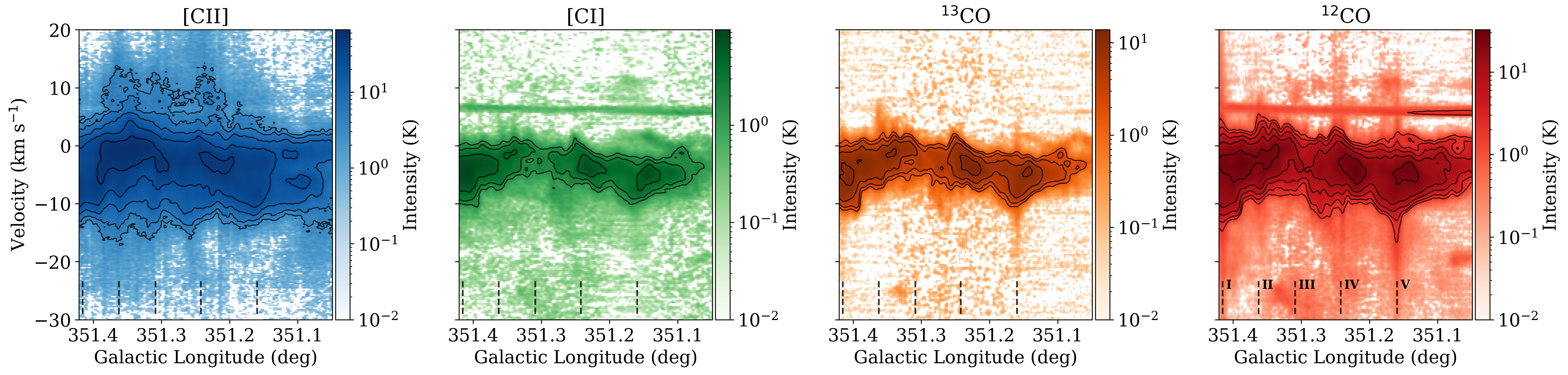}
    \caption{Position-Velocity (PV) maps of the four species toward the central filament. The slice is extracted along the Galactic longitude direction, centered at $b = 0.694^\circ$, spanning $l = 351.05^\circ$ to $351.42^\circ$, with a slice width of $\Delta b \approx 0.153^\circ$. Contour levels on each panel are at $3\sigma$, $5\sigma$, $10\sigma$, $20\sigma$, and $40\sigma$. Vertical black dotted lines mark the Galactic longitudes of FIR sources. }
    
    \label{lvplots}
\end{figure*}
\subsection{Emission morphologies of CO, \CI and \CII}\label{subsec:emissionmophology}

To illustrate the PDR-dominated nature of the region, three color composite maps of \cCI (blue), \cCII (green), and dust emissions (red) are presented in in Figure \ref{fig:3color_with_8_70mic}. While 8 $\mu$m traces PAH molecules excited by FUV radiation on cloud surfaces, 70 $\mu$m emission highlights warm dust associated with internal heating in PDR interfaces and denser regions. The widespread \cCII is strongly correlated with both dust tracers on logarithmic scale (Pearson coefficient r = 0.92 for 8 $\mu$m and r = 0.90 for 70 $\mu$m, see Fig. \ref{fig:cii_vs_dust}), confirming that it primarily arises from warm, FUV-illuminated atomic gas in the PDRs. Bright 8 $\mu$m emission appears as arc and rims toward the \HII regions, whereas bright 70 $\mu$m emission follows the filamentary ridge. \cCI shows only moderate correlation with the dust tracers (Pearson coefficient r = 0.36 for 8 $\mu$m and r = 0.53 for 70 $\mu$m), but stands out near FIR-I and in the filamentary structure extending north-south toward FIR-V. The emission morphologies of all four species are examined in detail in the following. 

We present moment 0 maps of the observed four species in Fig.~\ref{momentmaps}, separating the velocity ranges of $-$30 to $-$15 \kms, $-$15 to +5 \kms and +5 to +30 \kms. The intermediate range of $-$15 to +5 \kms represents the systemic velocity, as most $^{13}$CO emission is found within this velocity range (\citealt{Neupane2024}). The other velocity ranges indicate high-velocity features that are clearly seen in the \CII and \CI lines, though not always covering the same areas, and marginally in the $^{12}$CO line (Fig.~\ref{fig:spectra}). 

The $^{12}$CO emission is found to trace both dense and diffuse gas along the filament while $^{13}$CO emission mostly outlines the dense ridge regions (c.f. \citealt{Neupane2024}). $^{12}$CO also outlines both, blue-shifted and red-shifted wing emissions, tracing outflowing gas toward some FIR sources.

Within the systemic velocity range, the \CI\ emission distribution closely follows that of $^{13}$CO. In the blue-shifted interval ($-$30 to $-$15 km s$^{-1}$), this correspondence weakens, and \CI\ traces more extended emission, particularly around the FIR sources III and IV and along the western edge of the map. The red-shifted interval shows less prominent extended \CI emission, which is again confined to the western map boundary. In both high-velocity regimes, no spatial correspondence with \CII emission is observed.  

On the other hand, \CII emission prevails in all three velocity ranges. However, in the blue velocity range ($-$30 to $-$15 \kms), \CII is tracing patchy emission morphologies in and around the \HII regions. Most \CII emission is seen within the systemic velocity ranges, though the morphologies are quite different in comparison to all other tracers. At redder velocities (+5 to +30 \kms), \CII shows extended emissions above the central filament where there are no known \HII regions, while it shows bright and clumpy emission in and around the \HII regions along the filament. We also identify some ring-like structures that become better visible in channel maps and position-velocity cuts shown in Figure \ref{fig:cii_channelmap}. As a preliminary interpretation, we propose that the \CII emission beyond the systemic velocities is tracing  infalling or expanding gas and large-scale flows as well as turbulent ionized gas due to feedback from OB stars and \HII regions.

Comparing \CI and $^{13}$CO emission further in the common region, we find that some seemingly extended regions exhibit \CI emission above the $3\sigma$ level, while $^{13}$CO is below 3$\sigma$ for the integrated intensity maps at the systemic velocity range $-$15 to +5 \kms, and in the  ratio maps in the wider velocity range $-$20 to +20 \kms. We present the maps of this region in Fig.~\ref{fig:ci_not_13co}. Certainly the emission excess in \CI is not due to contamination from the local emission around +7 \kms. Regions where \CI is detected but $^{13}$CO is below 3$\sigma$ level, despite strong $^{12}$CO emission, likely correspond to low density or moderately shielded gas or \CI is additionally probing a dynamical processed gas such as inflows or cloud-collisions. 

The critical density of the $^{13}$CO\, transition ($\sim 1{-}1.5 \times 10^4\,\mathrm{cm}^{-3}$ at $T_\mathrm{kin} = 20$\,K) is an order of magnitude higher than that of the \CI line ($\sim 0.79{-}1.0 \times 10^3\,\mathrm{cm}^{-3}$). The critical densities are derived following \cite{Mangum2015} where the ranges reflect consideration of the ortho- and para-H$_{2}$ collision rate coefficients from LAMDA database (\citealt{Schoier2005}). In gas with density below this threshold, $^{13}$CO level populations are sub-thermally excited, and the line is faint. 
Additionally, \CI is more abundant in atomic to molecular transition zones and remains emissive where $^{13}$CO is partially photodissociated or insufficiently excited \citep{Tielens1985}. 
The detection of \CI emission that is correlated with $^{12}$CO but not with $^{13}$CO indicates a diffuse component in which \CI has not yet been ionized and traces lower density quiescent gas that has not yet been converted into molecular gas in star-forming regions.

On the other hand, \CI excess may indicate dynamically processed gas, potentially probing converging flows and cloud–cloud collisions, where shock compression, turbulence and partial CO dissociation together enhance relative abundance of atomic carbon (e.g. \citealt{Clark2019, FukuiY2018CC, Beuther2020, Beuther2026}). To investigate the dynamical scenarios, we produced spectra averaged over 60$''$ of the four species toward selected positions (P1 to P6), where \CI emission is above $3\sigma$ while $^{13}$CO is below $3\sigma$, presented in the Fig.~\ref{fig:ci_not_13co}. We identify multiple velocity components; a clear component in $^{12}$CO and \CI is observed in the velocity range $-$20 to $-$25 \kms\ (P1, P3, P6). The $\sim-$9 \kms\ component is detected in 5 of 6 positions, with three of those (P4, P5, P6) showing detection in all tracers  except $^{13}$CO. In one position (P2) the $\sim-$9 \kms\ emission is seen in $^{12}$CO and \CI, while in another position (P3) it is seen only visible in \CII. The $\sim-$4 \kms component, corresponding to the main molecular cloud velocity, is clearly detected in all positions, together with $\sim-$9 \kms\ in three positions. The coexistence of multiple velocity components and \CI emission excess, with respect to the $^{13}$CO, indicates dynamical interaction zone, possibly tracing converging flows or cloud-cloud collisions.

In Fig.~\ref{ccontinuum}, we further plot a 1.3 GHz MeerKAT radio continuum map \citep{Goedhart2024} toward NGC 6334 and overlayed \CII emission contours. The radio continuum emissions correlate nicely with the \HII regions. Overall, the bright \CII emission (contour at 400 K \kms) also shows morphological similarity with the bright radio continuum emission region. This hints that the bright emission in \CII at least partly results from the feedback due to embedded \HII regions in the central filament. Fraction of \CII emission may arise from collision with electrons inside the \HII regions and not from the PDRs surrounding it. Deriving an accurate fraction of \CII emission from collision with electrons is not straightforward. Observations of both [N II] and \CII lines (e.g. \citealt{Oberst2006}) is one way to constrain this which is beyond the scope of this work. However, using the 1.3 GHz radio continuum emission we can make qualitative assessments. Following \cite{Ramaila2026} (their equation 2) and assuming an electron temperature of \mybold{10$^{4}$~K} and using the 1.3 GHz flux density within the \HII region radii (adopted from \citealt{Anderson2014}), we derive an rms electron density ranging from $\sim$600 to 2000 cm$^{-3}$ toward the \HII regions. \mybold{Consequently, \CII emission originating from within the \HII region is thermalized.} At these moderate densities, theoretical models predict 10 to 30\% contribution of electron collision in the total \CII emissions (\citealt{Abel2006}), lower values for higher electron density. Comparable relative fractional contribution of ionized gas has been observed toward Carina Nebula (27\%: \citealt{Oberst2006}), while $\sim$5--15\% toward M43 region and other sub-regions in the Orion complex (\citealt{Pabst2022}).

\subsection{Position-Velocity ($PV$) maps}\label{subsec:lvmaps}
To evaluate the velocity extent of the emission, we first constructed global position-velocity (PV) maps 
of the four species, as presented in Fig.~\ref{lvplots}. The PV maps clearly illustrate that \CII and $^{12}$CO exhibit the most extended velocity distributions, while \CI and $^{13}$CO are confined to narrower velocity ranges relative to the systemic velocity. However, \CI emission is clearly broader than $^{13}$CO. Notably, \CII displays prominent blue-shifted and red-shifted velocity extensions, whereas $^{12}$CO predominantly traces blue-shifted gas, including bipolar outflows emanating from dense sources (e.g., FIR sources IV and V) along the central ridge. In $^{12}$CO and \CI, we also observe local emission at +7 \kms, while it is seen in absorption in the \CII map. From these PV maps, together with the emission morphologies presented in Fig. \ref{momentmaps}, it becomes obvious that \CII emission is clearly distinct to the other tracers. Unlike many FEEDBACK survey targets (\citealt{sofiafeedback}), the NGC 6334 region does not contain a single, spherical and  dominant \HII region, but harbors multiple embedded \HII regions along the dense filament. We thus performed PV-cuts crossing the individual \HII regions (Fig.~\ref{pvmaps_different_angles}). We do not observe any clear elliptical \CII emission distribution in the PV maps that are characteristic of fast ($>$10 km s$^{-1}$) expanding \CII bubbles (e.g., \citet{Luisi2021}) or ring-like emission with low-velocity ($<2$ km s$^{-1}$)  expansion signatures \citep{Dannhauer2025}. The only example of a fragmented ring is seen for G351.170+00.704.
We will come back to this point in the next section. The $^{12}$CO PV-maps show some high-velocity features due to outflow emission from protostellar sources. This emission arises from molecular clumps along the line-of-sight towards the \HII regions.

\subsection{[C II] high-velocity (HV) emission and comparison to radio continuum}\label{hv_emission}
\begin{figure*}[htbp!]
\centering
    \includegraphics[width=0.99\linewidth]{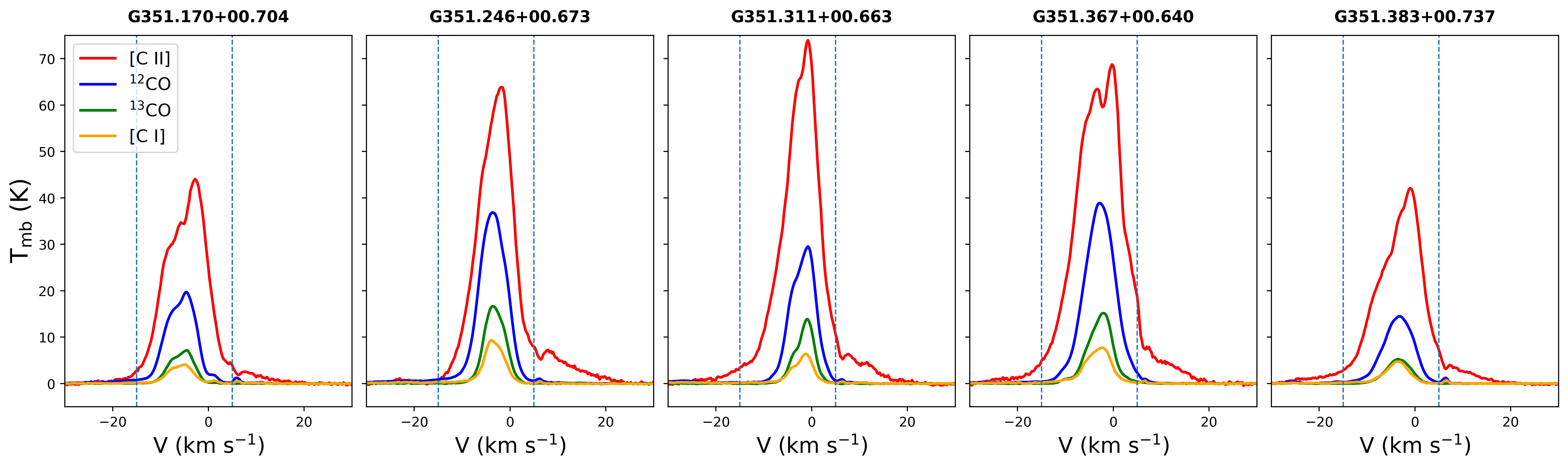}
    \caption{Average spectra toward \HII regions. Dotted blue lines indicate the systemic velocity range of $-$15 to +5 \kms\ toward NGC~6334.}\label{hiispec}
\end{figure*}

Average spectra toward the five \HII regions (Fig.~\ref{ccontinuum}), that are mostly covered in \cCII emission maps, are shown in Fig.~\ref{hiispec}. The line emission profiles, in particular in \cCII emission, reveal broad velocity ranges, with significant high-velocity (HV) components extending up to $\pm$30 \kms \, beyond the systemic velocity ($-$15 to +5 \kms) in most cases (see Fig.~\ref{hiispec}). All regions show HV red wing emissions. Except for G351.170+00.704 and G351.246+00.673, all regions also show HV blue wing emissions. The presence of HV wing emission \cCII in the average spectra is not found for the other species, even with $^{12}$CO emission which is known to trace HV molecular outflows. This alone already indicates that \cCII is tracing highly dynamic ionized gas toward these \HII regions.

\begin{figure*}[htbp!]
\centering    \includegraphics[width=0.99\linewidth]{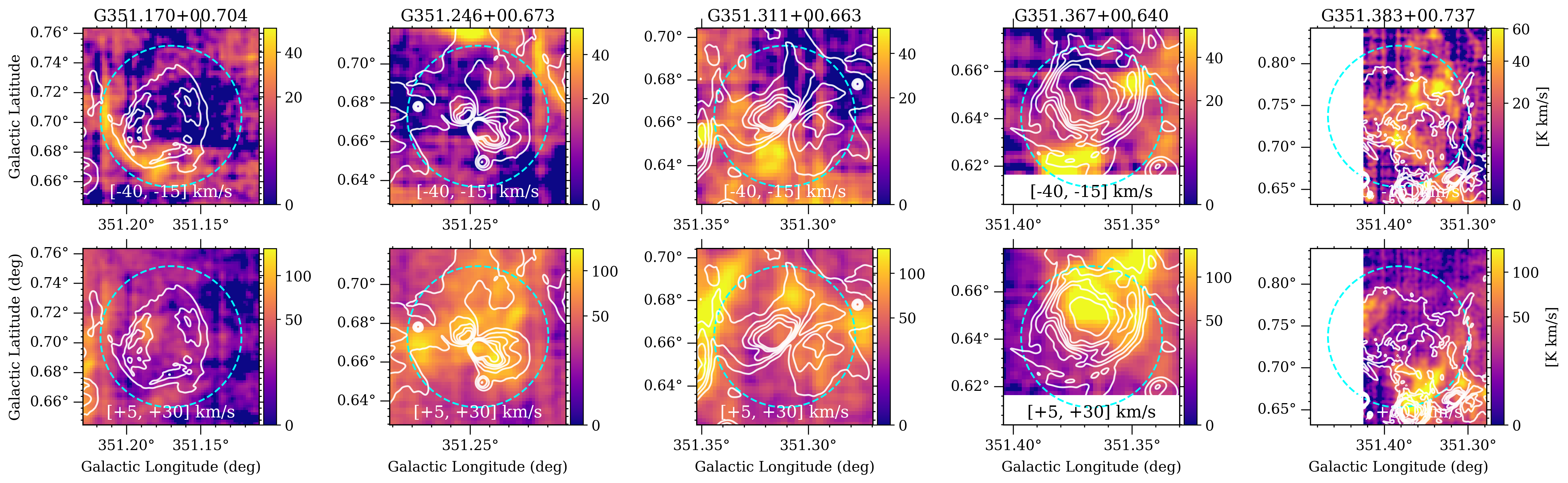}\\ 
    \caption{Maps of HV blue ($-$40 to $-$15 \kms) and red (+5 to +30 \kms) wing emission in \CII toward \HII regions (top and bottom panels, respectively). The radio continuum contours in white color are overlayed on both red and  blue emission maps at contour levels [0.01, 0.03, 0.05, 0.075, 0.1, 0.15] Jy/beam. The cyan dashed circles indicate the \HII regions, names labeled on top panels. }
    \label{hiiwingmaps}
\end{figure*}
To investigate the spatial distribution of the HV gas traced by \cCII, we constructed moment-0 maps for the blue ($-$40 to $-$15 \kms) and red (+5 to +30 \kms) wing emissions (Fig.~\ref{hiiwingmaps}). We overlaid the contours of the MeerKAT 1.3 GHz radio continuum emission on top of these maps. The emission morphologies reveal varying patterns across the five \HII regions that we briefly summarize in the following. 

G351.170+00.704 exhibits filled radio emission without compact structure. The blue HV emission matches arc-like emission along the edge of the \HII region, at least on one side. The red HV emission is filled and diffuse, but brighter than the blue HV emission.

G351.246+00.673 also reveals a bipolar structure in radio continuum (though more narrow compared to G351.311+00.663). The HV blue wing emission in \cCII is faint and filled, with some signatures of extended emission beyond the \HII region radii. Smaller, clumpy emission in blue seems to correlate with the bipolar emission seen in radio. The HV red wing emission is mostly filled within the \HII region radius, with diffuse extensions perpendicular to the bipolar radio emission.

G351.311+00.663 exhibits bipolar radio continuum emission. The HV blue emission in \cCII is more diffuse and filled, while red emission shows bright arc-like emission at the edge of the \HII region radius. This hints at an asymmetric PDR structure in red-shifted gas toward the \HII region.

G351.367+00.640 displays bright and filled radio emission. The blue HV emission shows two compact bright spots inside the edges of the \HII region radii. The red HV emission is clumpy and brighter, possibly indicating a partial shell of the \HII region, and extends outside the radius.

G351.383+00.737 exhibits diffuse and irregular radio continuum emission. Blue HV emission in \cCII is irregular, faint, and mostly diffuse. Red HV emission is relatively brighter (particularly around the bottom edge of the \HII region) and exhibits diffuse enhancements beyond the radius and corresponds well with the compact radio continuum structures.

Summarizing, we find hints of arc-like HV blue emission only toward the \HII region G351.170+00.704 and red HV emission is generally brighter than the blue one. In G351.311+00.663 and G351.367+00.640, HV red emission forms irregular or ring-like structures, tracing the edges of the \HII regions. Overall, \cCII emission does not seem to trace coherent ring-like structures or expanding bubbles associated with the \HII regions as was seen in many other studies \citep{Luisi2021, Tiwari2022, Bonne2022, Keilmann2025, Dannhauer2025} and as we already outlined in Sect.~\ref{subsec:lvmaps} for the PV-maps. Instead, it traces asymmetric and irregular diffuse enhancements, predominantly in HV red wing emissions, suggesting complex geometry and feedback interactions rather than coherent symmetric expansions.

The finding that HV \cCII emission is not always correlated with expanding bubbles or fragmented rings \citep{Dannhauer2025}, that are driven by the radiation and winds of the central OB-stars of the \HII regions, points toward a different interpretation of the emission. As proposed by \citet{Bonne2023}, it may indicate molecular cloud dispersal, in which the PDR surfaces of the associated molecular clouds are eroded by stellar feedback and give rise to HV \cCII emission. The molecular cloud is clearly highly filamentary and fragmented \citep{Arzoumanian2022, Neupane2024, Andre2025}, so that holes and channels allow radiation and winds to impact the cloud.

\section{Analysis}\label{sec:analysis}
In this section, we investigate the spatial structure of the emissions of different forms of carbon, perform quantitative analysis of velocity components, and analyze the line ratios from the integrated intensities. For the analysis based on multi-Gaussian fitting and line ratios we only use the common region where all data is available (box size 0.37$^{\circ} \times 0.15^{\circ}$ centered at $l=351.238$ and $b=0.69$, see Fig.~\ref{fig:mom0map1}). To understand the correlation with the dust properties we make use of cold dust emission in ATLASGAL 870 $\mu$m maps and warm/hot emission from Spitzer 8 $\mu$m maps from the archive.

\subsection{Spatial structure analysis via constrained diffusion decomposition}\label{sec:spatialstructure}
To investigate the spatial structure of the emission from different gas tracers, we applied the constrained diffusion decomposition (CDD) model \citep{Guang-xingLi2022} to the moment-0 maps. CDD decomposes the integrated intensity maps by solving modified non-linear diffusion equation that separates structures into different scales (characteristic sizes $\sim$2$^{k}$ pixels) while preserving positivity and avoiding negative artifacts common in wavelet based decomposition methods. The cumulative flux recovery at each scale quantifies the contribution of structure of different sizes and the residual fraction indicates unresolved or truly diffuse emissions. 

We applied CDD to the moment-0 maps (pixel scales of 6$''$ or equivalent physical size of $\sim$ 0.049 pc) in two velocity ranges: a broader range covering the whole velocity space (-30 to +30 \kms) and a more narrow range of the systemic velocity (-15 to +5 \kms). The objective is to investigate if inclusion of high velocity emissions, such as outflows or expanding gas, gives a different outcome from the analysis. The results, visualized in Fig.~\ref{fig:cdd}, illustrate the flux recovery fraction at different scales ($2^k$ pixels, $k = 0$ to 5). Table \ref{tab:cdd_results} summarizes the results.

\begin{table}[htbp]
\centering
\caption{Flux recovery scales (50\%) for different tracers.}
\label{tab:cdd_results}
\begin{tabular}{lcccc}
\hline
 & \multicolumn{4}{c}{Flux Recovery }  \\\cline{2-5}

 & \multicolumn{2}{c}{[$-$30, +30] km s$^{-1}$} & \multicolumn{2}{c}{[$-$15, +5] km s$^{-1}$} \\
\cline{2-5}
Tracer & \multicolumn{1}{c}{k (pixels)} & \multicolumn{1}{c}{Scale (pc)} & \multicolumn{1}{c}{k (pixels)} & \multicolumn{1}{c}{Scale (pc)} \\
\hline
$^{13}$CO & 2.82 & $\sim$0.35 & 2.89 & $\sim$0.37 \\
$^{12}$CO & 3.42 & $\sim$0.53 & 3.36 & $\sim$0.51 \\
\cCI  & 3.56 & $\sim$0.58 & 3.48 & $\sim$0.55 \\
\cCII & 4.87 & $\sim$1.47 & 4.96 & $\sim$1.53 \\
\hline
\end{tabular}
\end{table}

The molecular tracers $^{12}$CO, $^{13}$CO and \cCI consistently exhibit compact structures on $\sim$0.35--0.58 pc with residual fractions ranging from $\sim$8--29\%, indicating small variation between different velocity ranges. This illustrates that their emission is predominantly coming from dense cores and envelopes. In contrast, \cCII requires $\sim$1.5 pc scales for 50\% flux recovery, with high residual fractions ($\sim$48--49\%) in both velocity ranges. The CDD model of broader emission ($-$30 to +30 \kms) is only marginally different with respect to the narrower velocity range ($-$15 to +5 \kms), demonstrating that \CII traces extended, diffuse emission on cloud wide scales, even at systemic velocities, reflecting widespread UV illumination throughout the complex.

\subsection{Multi-Gaussian decomposition of \CII, \CI and $^{13}$CO}
We performed multi-Gaussian fitting to the \cCII, $^{13}$CO, and \cCI spectral lines using \texttt{gausspy+} algorithm (\citealt{Gausspyplus2019}) to derive the central velocity V (\kms), the velocity dispersion $\Delta$V (\kms) and  the amplitude T$_{\text{mb}}$ (K). Table \ref{tab:general_stats_threepanels} presents a summary of the statistics from the Gaussian fit results toward the common observational region and Fig.~\ref{fig:velpdf_and_lwdist}\subref{fig:sub_a} presents velocity probability density functions (PDFs) of the Gaussian components for each species.  The velocity PDF  clearly illustrates that \CII spans a wider velocity range (standard deviation of 5.45 \kms) than \cCI or $^{13}$CO, though their centroid velocities are comparable. 

\cCII exhibits the broadest velocity dispersions and highest mean brightness temperatures. The broad  dispersion, even from the Gaussian fitted individual components, suggests that it is tracing turbulently mixed ionized gas or directed flows, arising from photoevaporation of gas from PDR clump surfaces or wind-driven gas dynamic. In contrast, $^{13}$CO and \cCI show narrower line profiles with generally negative velocity skewness, consistent with denser, comparatively quiescent molecular gas possibly also affected by infall motions. Elevated kurtosis values for all three tracers point to multiple overlapping components and complex kinematics within the NGC~6334 central ridge.

Inside the bright \cCII region (T$_{\rm mb} \geq 55$,K, Fig.~\ref{fig:ciitmbpeak}), \cCII components are stronger and often require multiple Gaussian per line of sight, supporting moderate optical thickness or multiple velocity layers along the line of sight. $^{13}$CO and \cCI lines in this region are also slightly brighter and narrower than the map average, reflecting denser, more compact molecular gas. Skewness and kurtosis are enhanced, especially for \cCII and $^{13}$CO, highlighting the increased complexity and turbulence within the central ridge. Outside the bright region (T$_{\rm mb} < 55$ K), all three lines show weaker T$_{\text{mb}}$ and somewhat simpler profiles. \cCII retains a similar velocity width but with fewer bright peaks, tracing more diffuse ionized gas, while $^{13}$CO and \cCI exhibit slightly broader but less skewed distributions, consistent with relatively quiescent molecular environments.

Comparison of linewidth distribution of \cCII toward the \HII regions and ambient cloud environment is presented in Fig. \ref{fig:lw_hii_vs_ambient}. The ambient region is chosen at $l=351.081^{\circ}$ and $b=0.745^{\circ}$ westward to the FIR-V, with dynamic size matching the size of each \HII regions. The bulk \cCII linewidth distribution toward \HII regions remains comparable to the ambient cloud environment, but \HII regions exhibit a systematically enhanced high-linewidth tail, indicating localized feedback-driven dynamical broadening at irradiated cloud interfaces.

\subsubsection{Flux recovery and kinematic association with $^{13}$CO peak components}\label{subsec:peakvelcomp}
To understand kinematic association of the different tracers, we analyzed the Gaussian fit parameters by selecting the peak $^{13}$CO component in each pixel and matching it with the nearest \cCI and \cCII components in velocity space. The aim here is to disentangle velocity coupled gas traced by each species. Using a strict velocity tolerance of $\pm$0.75\,\kms (3 channels), only a limited match was achieved, with flux recovery fractions of 75\% for $^{13}$CO, 44\% for \cCI, and 21\% for \cCII, calculated by comparing Gaussian area maps to moment-0 maps. To better capture emission dynamically coupled with molecular gas, we adopted a wider velocity tolerance of $\pm$1.5\,\kms (6 channels) and selected the brightest \cCI and \cCII components where multiple were present.

This broader matching improved both spatial coverage and flux recovery reaching 77\% for $^{13}$CO, 62\% for \cCI, and 44\% for \cCII. The line width distributions of the velocity coupled gas show that $^{13}$CO and \cCI share similar velocity dispersions (Fig.~\ref{fig:velpdf_and_lwdist}\subref{fig:sub_b}), while \cCII remains broader, consistent with the results from the full component analysis (Table~\ref{tab:general_stats_threepanels}). These trends confirm that \CI closely follows the molecular gas traced by $^{13}$CO, while \CII partially overlaps but exhibits more turbulent components. 

Even with the wider tolerance, \CII recovers a smaller fraction of the total flux (21\% to 44\%) and indicates that about one third of the \CII emission is kinematically linked to molecular gas. However, its lower recovery compared to \CI and $^{13}$CO and its broader line widths suggest that much of the \CII emission arises from more diffuse or extended ionized layers, consistent with the predictions from the CDD model (see \S\ref{sec:spatialstructure}). Optical depth effects may contribute but likely play a secondary role.

\subsection{Correlation analysis of the four species}
To investigate how the intensities of $^{12}$CO, $^{13}$CO, \cCII, and \cCI relate across NGC~6334, we conducted a pixel-by-pixel correlation analysis using their moment-0 maps (pairwise threshold above 3$\sigma$) toward the common region. This approach allows us to examine how closely the different tracers probe common gas phases or distinct environments. Though $^{12}$CO is optically thick for most of the map and \cCII is likely optically thick toward the brightest regions, we however perform a line ratio analysis from the integrated intensities. Accordingly, this demands caution in interpreting the results.

\begin{figure}[t!]
\centering
    \includegraphics[width=0.80\linewidth]{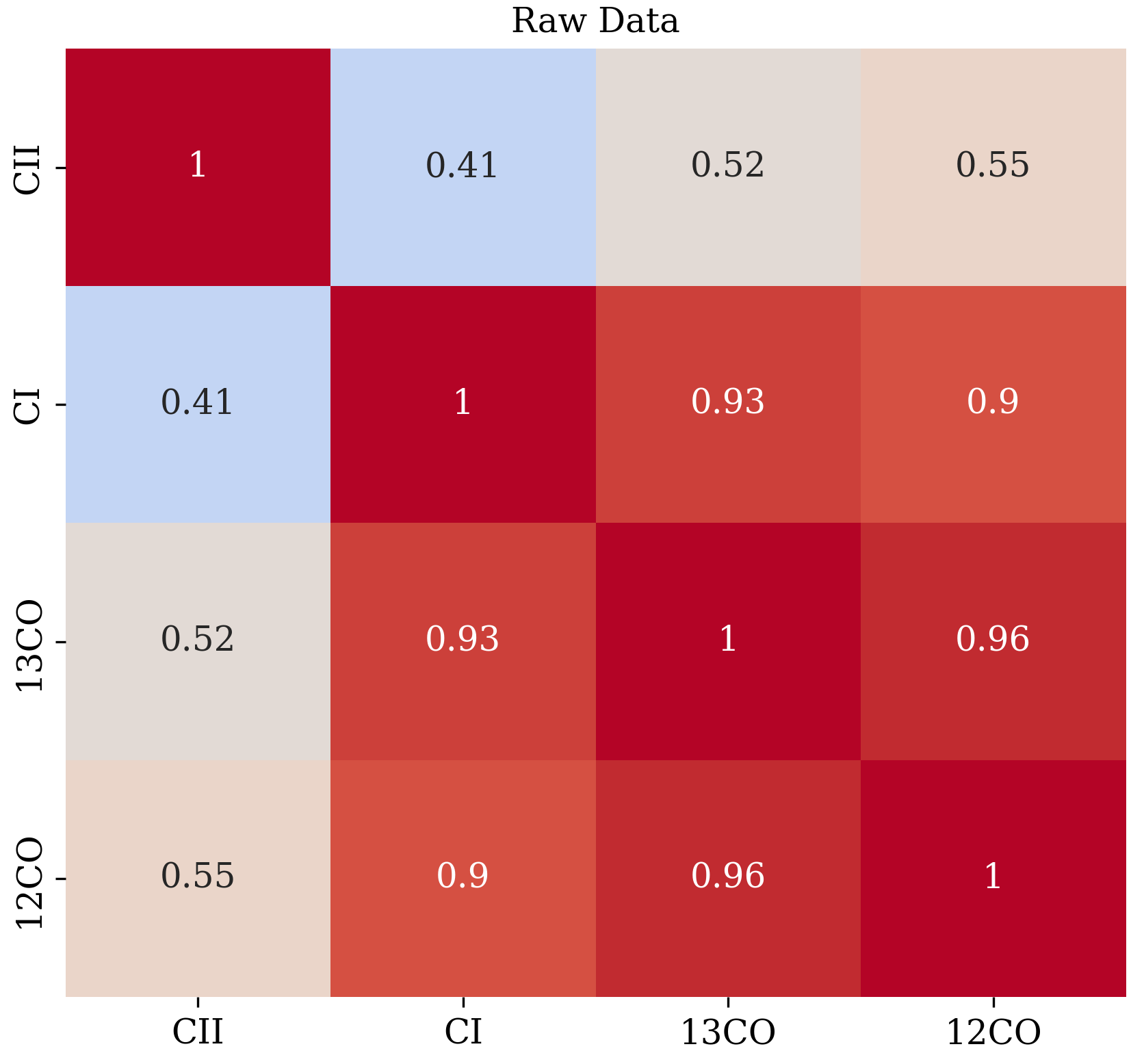}
    \caption{Heat map showing correlation matrices (Pearson coefficients) between the pair of species from moment-0 maps (T$_{\rm mb}$dV) toward the common region. }\label{heatmap_corr}
\end{figure}

\begin{figure*}[htp]
    \centering
    \includegraphics[width=0.31\linewidth]{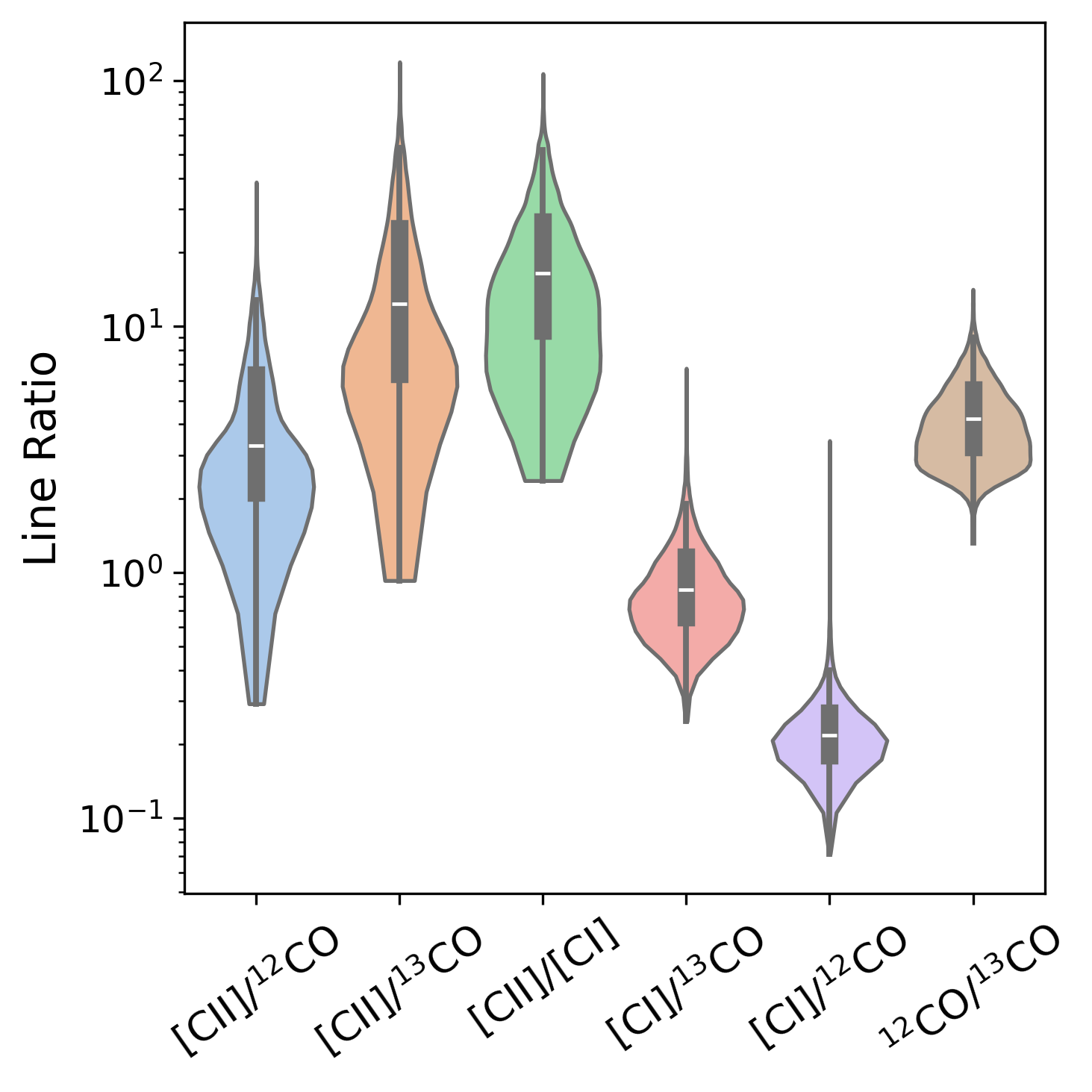}
    \includegraphics[width=0.31\linewidth]{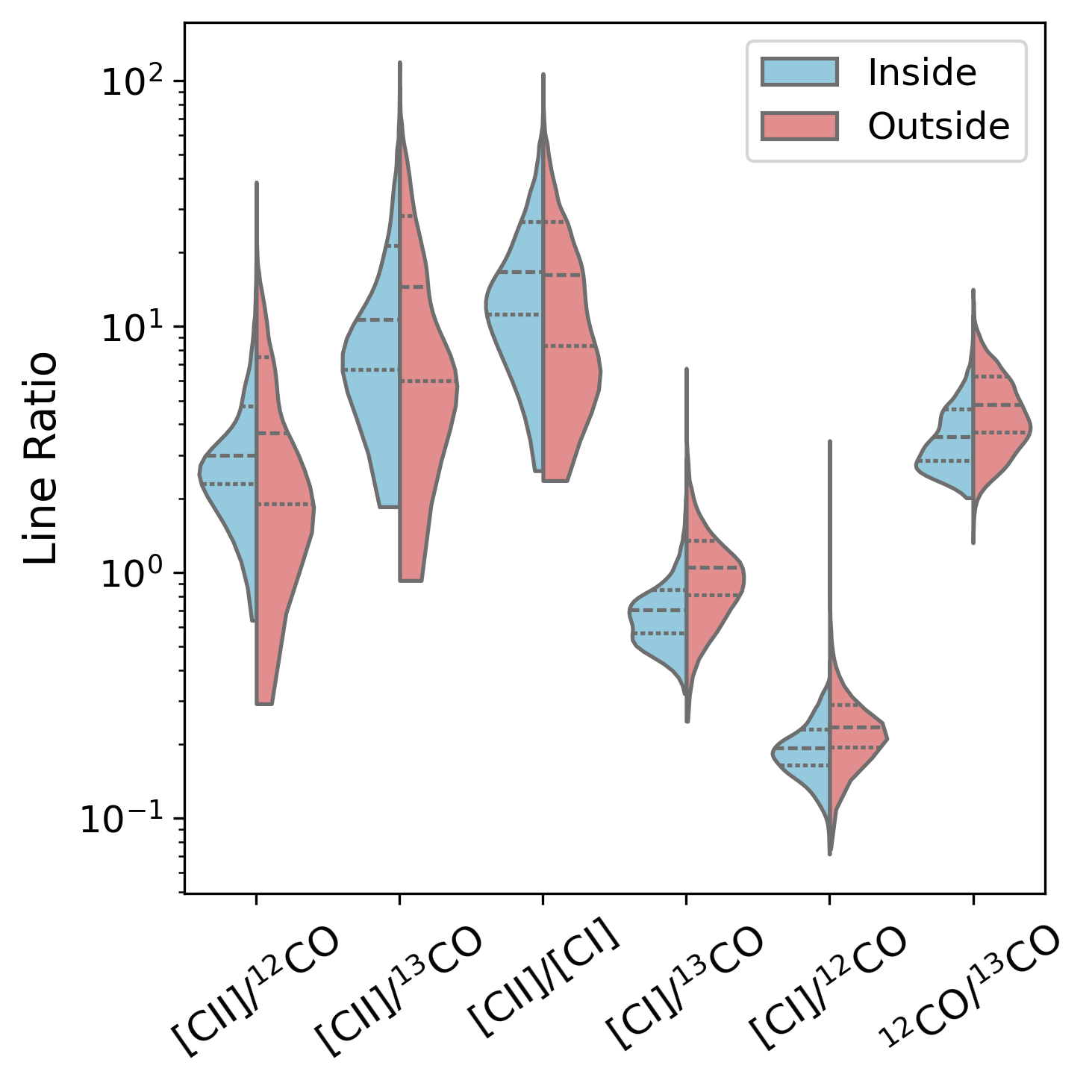}
    \includegraphics[width=0.32\linewidth]{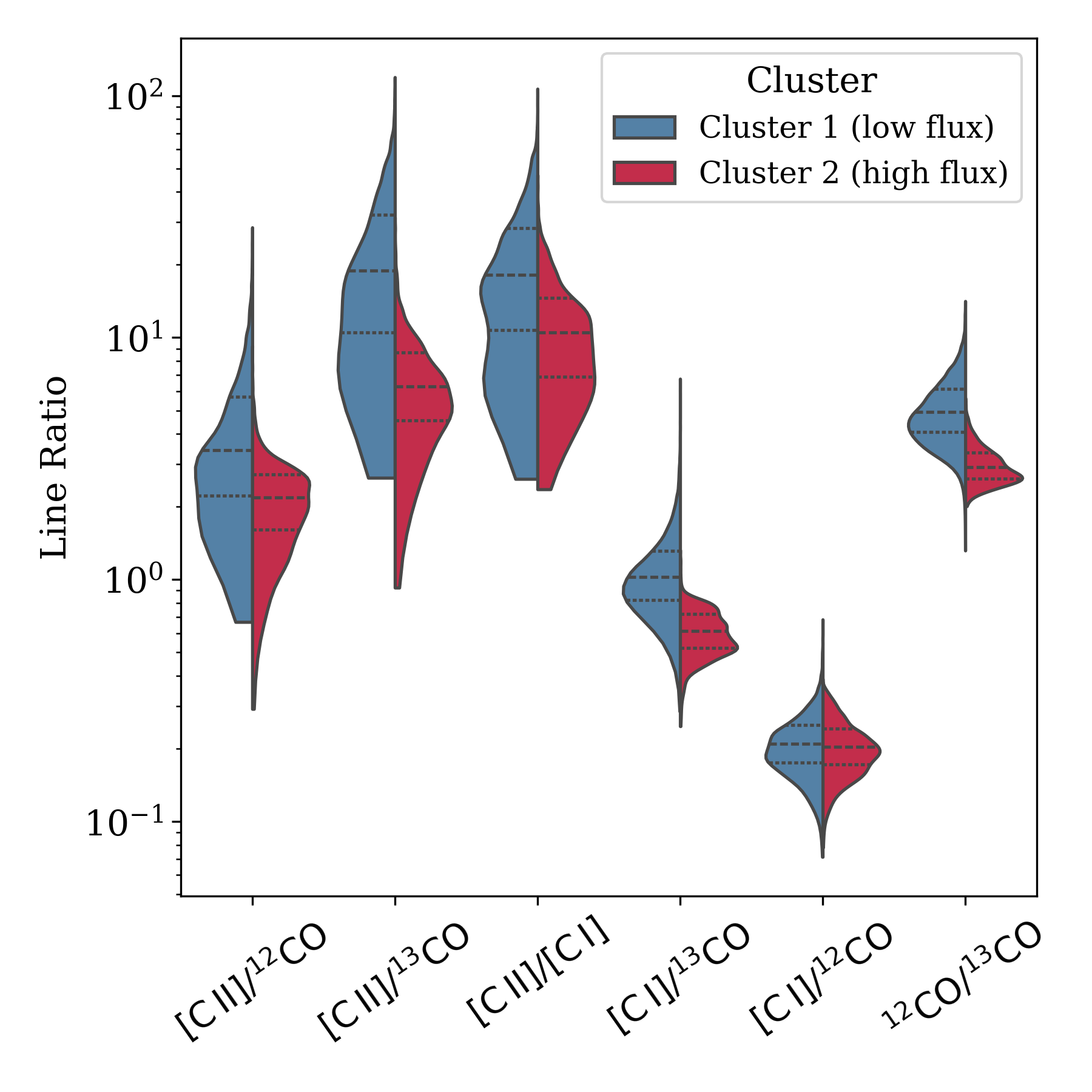}
    \caption{Line ratios of the species shown in violin plots. Left: full map, middle: inside vs outside, right: Cluster1 vs Cluster2. The black box and white horizontal line inside the violins (left panel) indicate the interquartile range and median value, respectively. In the split-violin plots (middle and right panels), the internal horizontal lines display the individual quartiles and medians for each group.} 
     \label{fig:lineratios}
\end{figure*}

We evaluated three cases: (1) the full map (i.e., the common mapped region), (2) the region inside the \cCII $T_{\rm mb} \geq 55$ K contour, and (3) the region outside of this contour (Fig.~\ref{fig:ciitmbpeak}). 

First, Pearson correlation coefficients were computed directly from the moment-0 maps, as shown in Fig.~\ref{heatmap_corr}. We also calculated correlations using standardized and arcsinh transformed data, which reduce the influence of heavy-tailed intensity distributions. The resulting coefficients differ only slightly, indicating that the observed correlations are robust against non-Gaussian intensity effects.

Across the full map, $^{12}$CO, $^{13}$CO, and \cCI show very strong correlations ($r=0.90-0.96$), reflecting their common origin in the molecular gas reservoir. \cCII exhibits more moderate correlations with $^{12}$CO ($r=0.55$) and $^{13}$CO ($r=0.52$), and a weaker link with \cCI ($r=0.41$).

Within the bright \cCII region ($T_{\rm mb} \geq 55$ K), the molecular tracers remain tightly correlated ($r \gtrsim 0.87$), while correlations between \CII and the molecular lines weaken ($r = 0.24$–$0.47$). This indicates that although \cCII emission is intense here, a significant fraction originates from gas not directly coupled to the dense molecular component, likely from PDR surfaces or partially ionized layers. This supports our earlier result that \cCII exhibits broad velocity dispersions and requires multiple Gaussian components to reproduce its line profiles. 

Outside the 55 K contour, molecular tracers remain extremely well correlated ($r = 0.91$–$0.96$), whereas \cCII becomes largely decoupled ($r = 0.13$–$0.21$). This contrast reinforces the dual nature of \CII emission: it partly traces dense PDR gas within the bright ridge but is increasingly dominated by diffuse ionized material in fainter regions. 
\subsection{Line ratios and physical conditions}\label{subsec:lineratios}
We analyzed the pixel-by-pixel line ratios of \cCII, \cCI, $^{12}$CO, and $^{13}$CO to explore spatial variations in the physical conditions. Ratios were computed from moment-0 maps for pixels above 3$\sigma$ in both species. Figure~\ref{fig:lineratios} shows the distributions of the line ratios, and Table~\ref{tab:line_ratios} summarizes their statistics. 

Across the common mapped region, ratios involving \cCII are systematically higher and more broadly distributed than those among the molecular tracers. The \cCII/$^{12}$CO and \cCII/$^{13}$CO ratios show extended high ratio tails, reflecting bright \cCII emission relative to CO in regions of diffuse or ionized gas. The \cCII/\cCI ratio is similarly large but with a more narrow spread, indicating partial spatial and physical overlap between the ionized and atomic carbon layers. Ratios between molecular tracers, particularly \cCI/$^{13}$CO and $^{12}$CO/$^{13}$CO, remain more uniform, consistent with their shared molecular origin. 

We note that opacity correction in \cCII can only increase the ratio values in \cCII/$^{13}$CO and \cCII/\cCI. Since both, \cCI and $^{13}$CO are likely optically thin, opacity effects in these ratios are negligible \mybold{to minor}. Considering only velocity coupled components, the overall trends derived from moment zero based line ratios are preserved (Table \ref{tab:ratio_stats_peakcomp_unclipped}), including the relative ordering of the ratios and their positive skewness. Thus, the line ratio analysis from the integrated intensities is valid to understand their general trends in dense and diffuse regions while derivation of physical parameters based on these may lead to larger uncertainties. 

\subsubsection{Inside vs outside the \CII bright region}
To examine environmental variations, we investigate in the following regions inside and outside the \cCII intensity contour at $T_{\rm mb} = 55$ K. 
Inside the $55$ K contour, the mean \cCII/$^{12}$CO and \cCII/$^{13}$CO ratios ($\sim4$ and $\sim17$, respectively) are slightly lower than the full map averages, while molecular ratios such as \cCI/$^{13}$CO and $^{12}$CO/$^{13}$CO remain nearly constant ($\sim0.8$ and $\sim3.9$) (see Table~\ref{tab:line_ratios}). This indicates that, although \cCII is bright, its emission here arises from a mixture of dense and diffuse UV illuminated PDR gas. 

Outside the contour, mean \cCII ratios increase modestly (\cCII/$^{12}$CO $\sim5$, \cCII/$^{13}$CO $\sim19$), reflecting a higher relative contribution of diffuse \cCII to fainter CO. In contrast, \cCI/$^{13}$CO rises from $\sim0.8$ inside to $\sim1.1$ outside, while $^{12}$CO/$^{13}$CO grows from $\sim3.9$ to $\sim5.1$. These trends point to enhanced atomic carbon in the more diffuse outer regions, consistent with the morphological trends discussed in Section~ \ref{subsec:emissionmophology}.

\begin{figure*}[htbp!]
    \sidecaption
    \includegraphics[width=12cm]{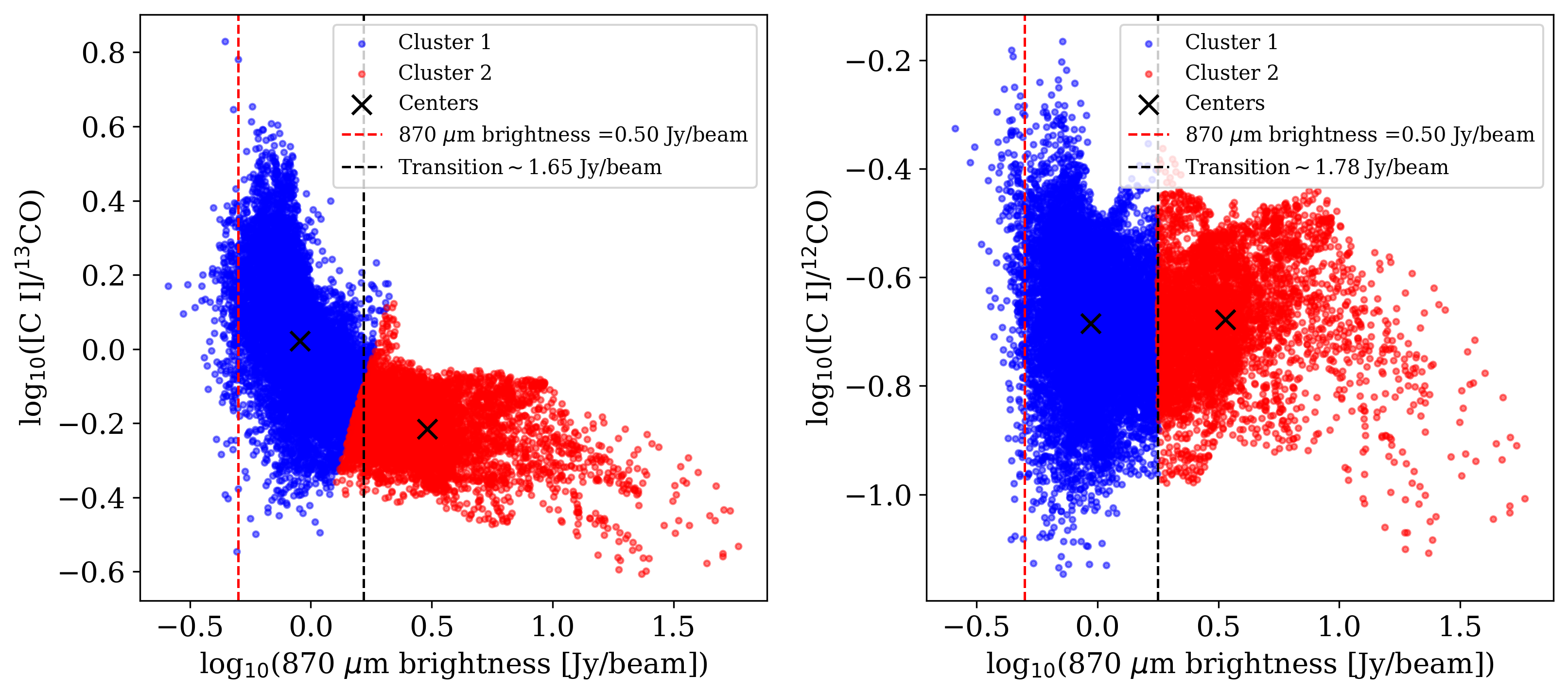} 
    \caption{\cCI/$^{13}$CO and \cCI/$^{12}$CO ratio as a function of ATLASGAL 870 $\mu$m brightness in a log-log scale. The blue and red color data points are Cluster~1 and Cluster~2 from K-means clustering. The black crosses represent mean positions of each clusters.}    
    \label{fig:ci2co13_vs_atlasgal}
\end{figure*}

\subsubsection{K-Means clustering with ATLASGAL surface brightness}
To complement the contour based analysis, we examined the \cCI/$^{13}$CO ratio as a function of ATLASGAL 870 $\mu$m surface brightness (Fig.~\ref{fig:ci2co13_vs_atlasgal}, left panel). Using pixels above $3\sigma$ in both lines, we applied K-means clustering (using the
scikit-learn package \citep{scikit-learn}), which separated the data into two populations: a low-brightness cluster (Cluster~1) and a high-brightness cluster (Cluster~2), with 870 $\mu$m brightness transition near 1.6–1.8\,Jy\,beam$^{-1}$. 

Cluster~1 shows systematically higher ratios involving both \cCII and \cCI (e.g., \cCII/$^{13}$CO $\sim23$, \cCI/$^{13}$CO $\sim1.1$) (Table~\ref{tab:kmeans_stats}), suggesting that carbon is more readily in ionized or atomic form in lower density and more UV exposed environments. Cluster~2, coinciding with the bright \cCII region, exhibits lower ratios (\cCII/$^{13}$CO $\sim7$, \cCI/$^{13}$CO $\sim0.6$) characteristic of predominantly molecular gas. Morphologically, the contour of \cCI/$^{13}$CO $\approx1$ closely follows the $T_{\rm peak}$(\cCII)$=55$ K boundary, except toward the western ridge near FIR~V (Fig. \ref{fig:clustermasks}).

\begin{figure}[htbp!]
    \centering
    \includegraphics[width=0.75\linewidth]{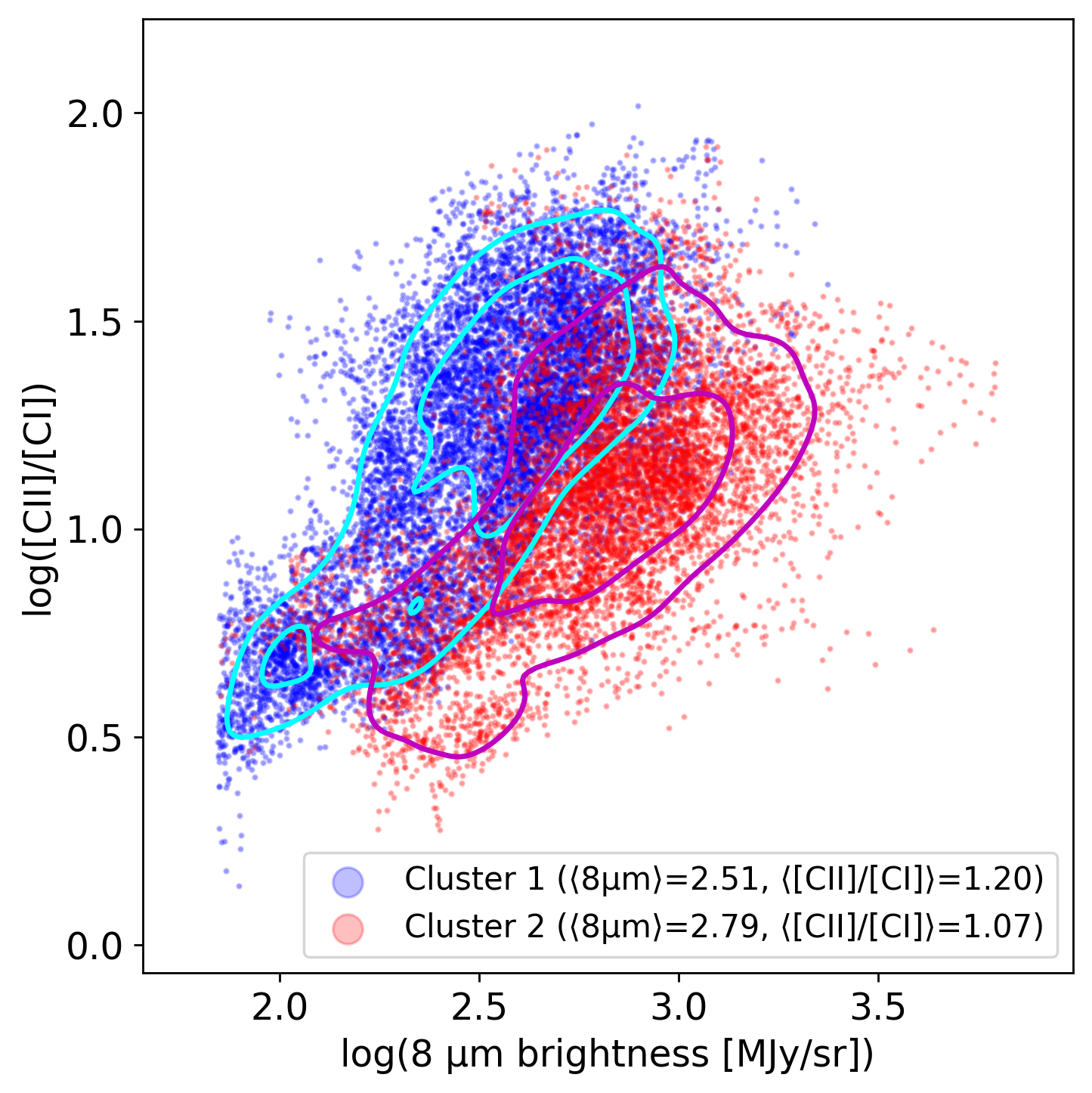}
    \caption{\cCII to \cCI ratios with respect to the 8 $\mu$m flux. The blue and red color data points represent Cluster~1 (low ATLASGAL brightness) and Cluster~2 (high ATLASGAL brightness), respectively. Cyan and magenta contours show kernel density estimate (KDE) iso-density levels for the two clusters.}
    \label{fig:cii_to_ci_vs_8mic}
\end{figure}

We investigated the \cCII to \cCI ratio of the two clusters versus the 8 $\mu$m brightness, presented in Fig.\ref{fig:cii_to_ci_vs_8mic}. While the mean value of 8 $\mu$m flux for Cluster~1 is about half the value of Cluster~2, \cCII to \cCI ratio is systematically higher in Cluster~1 for ranges of 8 $\mu$m brightness. This suggests diffuse regions favor \cCII dominated warm PDR layers with weaker PAH excitation while denser regions show relative dominance of \cCI despite stronger PAH emission from the exteriors revealing density stratified chemical structure. 

We further performed similar clustering analysis of the \cCI to $^{13}$CO ratio with respect to the H$_{2}$ column density (from PPMAP, \citealt{Marsh2015}) and 70 $\mu$m emission (\citealt{Carey2009}). It turns out that the transitional column density between the high-brightness and low-brightness clusters is approximately at 1.7$\times$10$^{22}$ cm$^{-2}$ ($A_{\rm V}\sim 18.6$; derived following \cite{Bohlin1978}), a consistent result from all four methods investigated here (Fig. \ref{fig:clustermasks}). At this transitional layer, the corresponding  dust temperature is approximately 28 K (also from PPMAP, \citealt{Marsh2015}). This depth is well within the dense gravity dominated column density regime in molecular clouds (e.g., \citealt{Schneider2015}). These results imply that a significant fraction of carbon may reside in “CO-dark” molecular gas in NGC~6334. \mybold{Models of molecular cloud formation via colliding flows also predict that “CO-dark" gas may dominate the early gas for substantial fraction of their assembly history (e.g., \citealt{Clark2012})}. Additionally, the deep penetration of the UV field may suggest a porous and inhomogeneous cloud geometry. 

\section{Discussion}
\label{sec:discussion}
Most of PDR theories and models (e.g., \citealt{Tielens1985,LePetit2006,Roellig2022,Pound2023}) consider a one dimensional layered geometry for UV-irradiated molecular clouds in which the spatial distribution of \cCII, \cCI, and CO emission are interpreted. These tracers are expected to probe different chemical and thermal layers as an effect of progressive attenuation of the radiation field into the cloud interior. In this simplified picture, C$^+$ is expected to dominate the outermost layer, C$^0$ traces the intermediate layer, and CO is found in the denser interior part. The \CII emitting surface layers arise from regions of $A_V \lesssim 1$–3, while [C I] and CO originate from the regions where the gas becomes shielded from the incident FUV radiation. However, molecular clouds exhibit more complex geometry, and assessing the validity of this stratified picture requires multi-wavelength data sets at high angular and velocity resolution.  In this work, we obtained state of art data set for each species toward NGC 6334 central region. We discuss our findings in the following.

\subsection{Stratification of the carbon phases in NGC~6334}
The morphology of the emission distribution (Fig. \ref{momentmaps}) highlights clear differences among the tracers (see Sect.~\ref{subsec:emissionmophology}). \CII is widespread and extends far beyond the dense ridge, while $^{13}$CO is compact and confined to the dense ridges. $^{12}$CO and \CI occupy an intermediate regime, tracing both dense regions and surrounding envelopes. This hierarchical distribution is further supported by the spatial scale analysis (Sect.~\ref{sec:spatialstructure}), where both CO, and \CI recover a significant fraction of their emission on scales of $\sim$0.3–0.5\,pc, characteristic of dense clumps and filamentary interfaces, whereas \CII remains dominated by larger-scale emission. Thus, at first order, the overall large scale picture is consistent with classical stratified carbon phases, in which carbon transitions from ionized to atomic to molecular form with increasing shielding and density.

Looking into the details, we observe the spatial distributions of the tracers with significant overlap, with \CII extending far beyond classical surface layers and \CI bridging both diffuse and dense PDR gas components. A significant fraction of the \CII emission arises from large scale ($\geq$ 1 pc), diffuse PDR gas that is dynamically and spatially decoupled from the molecular phase, as only $\sim$20–40\% of the \CII\ emission is velocity coupled to the dense $^{13}$CO components (Sect.~\ref{subsec:peakvelcomp}). The positive correlation with radio continuum emission and \CII bright regions associated with the \HII regions suggest that significant \CII emission arises from local feedback. These results together indicate that, while a global stratification of carbon phases is preserved, the carbon phases are also locally mixed and distributed within NGC~6334 by density inhomogeneities, stellar feedback, and possible inflows or cloud-cloud collision.
\subsection{Kinematic signatures and dynamical (de-)coupling}
The velocity structure of the carbon tracers in NGC~6334 reveals a clear dynamical differentiation among the phases (see Sect.~\ref{subsec:lvmaps}). \CII exhibits the broadest velocity spread and often requires multiple components to fit its line profiles, indicative of multi-layered, turbulent, inflows or feedback-processed gas. $^{12}$CO shows second broadest emission, including high-velocity wings associated with some of the FIR sources located in the dense filamentary ridge. In contrast, $^{13}$CO remains narrow and confined, and \CI displays an intermediate behavior, sometimes showing blue-shifted emission not present in $^{13}$CO but seen in $^{12}$CO. Median linewidth of \CI in logarithmic ATLASGAL 870~$\mu$m  bins show larger values in comparison to $^{13}$CO for brightness bins up to $\lesssim$5 Jy/beam, that is more prominently observed in the 95th percentile (P95) values. At higher brightness bins both tracers show comparable median linewidth while $^{13}$CO exhibits broader P95 values (see Fig. \ref{fig:ci_13co_linewidth}). \mybold{The linewidth distribution presented in Fig. \ref{fig:ci_13co_linewidth_violin} further illustrate the trend for outside vs inside regions.} The high-velocity \CII emission does not spatially match with the outflow features in the radio continuum, implying its minor role, whereas the positive association of \CII with overall radio continuum emission morphologies around \HII regions points to stellar feedback as the dominant driver. Gaussian decomposition confirms that \CII has significantly broader line widths than $^{13}$CO and \CI, even when only velocity-coupled components are considered, reflecting a combination of high turbulence and multiple unresolved line-of-sight layers, including the possible existence of converging flows, components related to cloud-cloud collisions, and components related to the cloud dispersal from the local feedback. Overall, these results indicate that \CII traces dynamically complex, diffuse PDR gas shaped by large scale feedback and flows, whereas CO and \CI probe progressively denser and quiescent structures, highlighting a pronounced dynamical decoupling between the ionized, atomic, and molecular carbon phases.
\subsection{Line ratios and physical stratification}
The line ratio and clustering analyses (Sect.~\ref{subsec:lineratios}) highlight the chemical differentiation across NGC~6334, complementing the spatial stratification revealed by the morphology. The \cCI/$^{13}$CO ratio increases outward from the dense ridge to the diffuse envelope, showing that atomic carbon extends beyond molecular ridges and clumps. Ratios involving \CII\ are systematically higher and positively skewed, with enhanced values outside the dense central ridge where diffuse emission dominates (see Figure \ref{fig:cii_to_ci_vs_8mic}). The transition in [\CI]/$^{13}$CO around H$_2$ column densities of $\sim1.7\times10^{22}$\,cm$^{-2}$ ($A_V \sim 18$) indicates that FUV photons penetrate deep into the cloud along cavities and clump interfaces, enhancing atomic carbon abundance in the PDR interiors (Sect.~\ref{subsec:lineratios}). Taken together with the spatial and kinematic results, these chemical diagnostics reinforce a coherent picture of NGC~6334 as a globally stratified but locally complex medium. The stratification is preserved on large scales, while turbulence, flows, stellar feedback, and cloud porosity mix the layers locally, producing chemically distinguishable but dynamically (de-)coupled phases.

\subsection{[CI] and [CII] as dynamical tracer of flows} 
In addition to the stratified and chemically differentiated structure, the carbon tracers reveal signatures of dynamically interacting gas in NGC~6334 (see \S~\ref{subsec:emissionmophology}). We observe regions that exhibit \CI emission above $3\sigma$ where $^{13}$CO is below the noise level (Fig. \ref{fig:ci_not_13co}), despite strong $^{12}$CO emission. These could indicate low density or moderately shielded gas in which \CI remains emissive while $^{13}$CO is sub-thermally excited or partially photodissociated. However, we find multiple velocity components in the averaged spectra (see Fig.~\ref{fig:ci_not_13co}), particularly around $-20$ to $-25$ \kms and $-9$ \kms in addition to the systemic velocity of $\sim-4$ \kms, suggesting overlapping gas velocity components. In some regions, the $-9$\,\kms component is visible only in \CII, while in most regions it is clearly identifiable in $^{12}$CO and \CI. 

This highlights that \CI and \CII can also be tracing atomic layers that are dynamically distinct. Such features are consistent with predictions of converging flows or cloud–cloud collisions, where shock compression, turbulence, and partial CO dissociation together enhance the atomic carbon abundance \citep{Clark2012,Clark2019,FukuiY2018CC}. NGC~6334 has been reported to be undergoing large scale cloud-cloud collision by \cite{FukuiY2018CC}. In the extended NGC~6334 filamentary complex (extent of $\sim$ 80~pc), we identified multiple velocity components ($[-25, -15]$ \kms\ feature only in $^{12}$CO and $-9.2$ and $-3.9$ \kms\ components in $^{13}$CO) and `bridge' features in velocity ranges $-12$ to $-7$ \kms) associated with the cloud-cloud collision (\citealt{Neupane2024}). The collision time scale is about 0.4 Myr, considering 10 pc spatial scale of the central filament (c.f. \citealt{FukuiY2018CC}). Time dependent chemical models suggest that full conversion of C{\sc{i}} to CO requires time scale of the order of Myr \mybold{(e.g., \citealt{Suzuki1992,Aikawa2026})}. In young cloud like NGC~6334, the dynamical interaction can dominate the observed chemical behavior of the different carbon phases, particularly toward the interaction regions. 
Together, these observations imply that a fraction of the \CI and \CII emission probes dynamically processed regions in which dense and diffuse layers interact. Thus these tracers can also be effective probes of kinematic features in converging flows or cloud–cloud collisions. 

\section{Summary and conclusions}\label{sec:summary}

We used high-resolution observations of the three primary carbon phases, ionized carbon (C$^+$), neutral carbon (C$^0$), and carbon monoxide ($^{12}$CO and $^{13}$CO 3$\to$2), to investigate the multi-phase structure of the NGC~6334 central ridge and its environment. By combining morphological, kinematic, spatial scale, and line ratio analyses, we examined how ionized, atomic, and molecular gas coexist and interact under the influence of strong stellar feedback in the region.

\begin{itemize}

\item Our observations show that ionized, atomic and molecular gas co-exist in the NGC~6334 central ridge, however, they are only partially coupled. \CII is most extended and spread beyond the dense molecular ridge, tracing both warm PDR surfaces (well correlated with 8 and 70 $\mu$m continuum) and feedback from H II regions (correlated with radio continuum). $^{13}$CO is compact and confined to the dense ridge while $^{12}$CO and \cCI occupy an intermediate regime. $^{12}$CO additionally traces outflowing material linked to protostellar activity toward FIR sources in the central ridge. 

\item \CII shows broadest velocity coverage that includes diffuse PDR to the feedback driven components around the \HII regions. Only 20-40\% of the \CII is velocity coupled to $^{13}$CO suggesting significant gas (de-)coupling. $^{13}$CO and \CI stay similar except that \CI is slightly extended in morphology more in bluer velocities with respect to systemic velocity and shows broader high line widths tail both in the dense and diffuse regions.

\item Spatially \CII  remains largely diffuse and does not reach total flux recovery of 50\% even at scales of $\sim$1.5 pc, while CO and \CI recover half their total flux on ~0.3–0.5 pc scales (clumps/filaments). These results indicate \CII traces predominantly the larger scale diffuse PDR gas while CO and \CI are associated with progressively denser, more localized regions. \HII regions contribute up to 30\% of the local \CII emission tracing the ionized gas.  

\item $^{13}$CO and \CI are generally described by narrower components, while \CII often exhibits multiple and broader components highlighting the kinematic complexity of the region. Comparison of \CII line widths toward \HII regions versus ambient medium further highlights the thermal plus dynamical broadening of the line due to local feedback.

\item The line ratios \CII/\CI and \CII/CO  are higher and skewed, signaling extended UV-irradiated diffuse gas. \CI/$^{13}$CO on the other hand shows a transition around H$_2$ column densities of $\sim1.7\times10^{22}$\,cm$^{-2}$ ($A_V \sim 18$) deeper into the cloud separating the dense ridge (moderate ratios) from the outer diffuse envelope (higher ratios $\gtrsim$1), indicating deep FUV penetration and a porous cloud geometry.

\item We observe \CI and \CII also tracing distinct velocity components as in CO ($\sim -$20, $-$9, and systemic $-$4 \kms) that are associated with cloud-cloud collisions and conclude that \CI and \CII also probe dynamically processed gas. 

\end{itemize}

Together, these results depict the NGC~6334 central ridge as a strongly layered, feedback influenced photon-dominated region. Dense molecular gas traced by $^{12}$CO, $^{13}$CO, and \CI\ is embedded within a widespread, dynamically complex envelope of diffuse \CII\ emission that is only partially coupled to the molecular gas. FUV radiation penetrates deeply into the molecular ridge, enhancing atomic and ionized carbon even at high column densities. This implies that a significant fraction of carbon may reside in `CO-dark' gas, shaped by feedback and the porous, clumpy cloud geometry. Overall, this study highlights that NGC~6334 is a complex, prototypical example of a clumpy, multi-phase PDR in which stellar feedback governs chemical layering, energy balance, and dynamical structure and evolution of the molecular cloud.

\begin{acknowledgements}
This study was based on observations made with the NASA/DLR  Stratospheric Observatory for Infrared Astronomy (SOFIA). SOFIA is
jointly operated by the Universities Space Research Association Inc. (USRA), under NASA contract NNA17BF53C and the Deutsches SOFIA
Institut (DSI), under DLR contract 50 OK 0901 to the University of Stuttgart. upGREAT is a development by the MPIfR and the KOSMA/University Cologne, in cooperation with the DLR Institut f\"ur Optische Sensorsysteme.
This work was carried out in part at the JPL, California Institute of Technology, under contract with 
NASA. \\
Financial support for FEEDBACK at the University of Maryland was provided by NASA through award SOF070077 issued by USRA. The FEEDBACK project was supported by the BMWI via DLR, project number 50OR2217. \\
This work was supported by the CRC 1601 (SFB 1601 sub-projects B1, B2) funded by the Deutsche Forschungsgemeinschaft (DFG) – 500700252 and by DYNAVERSE.
The CO and [CI] data were collected under the Atacama Pathfinder EXperiment (APEX) Project, led by the Max Planck Institute for Radio Astronomy. S.N. also acknowledges continuous support from the APEX staffs and telescope operators during the observations. G.G. gratefully acknowledges support by the ANID BASAL project FB210003. \\
This work made use of Astropy (\url{http://www.astropy.org}) a community-developed core Python package and an ecosystem of tools and resources for astronomy \citep{Astropy2022}. 

\end{acknowledgements}

\bibliographystyle{aa}
\bibliography{biblio}

\begin{appendix}
\onecolumn
\section{Additional plots}\label{app-sec-a}
\FloatBarrier
\begin{figure*}[htbp!]
    \centering
    \includegraphics[width=0.65\linewidth]{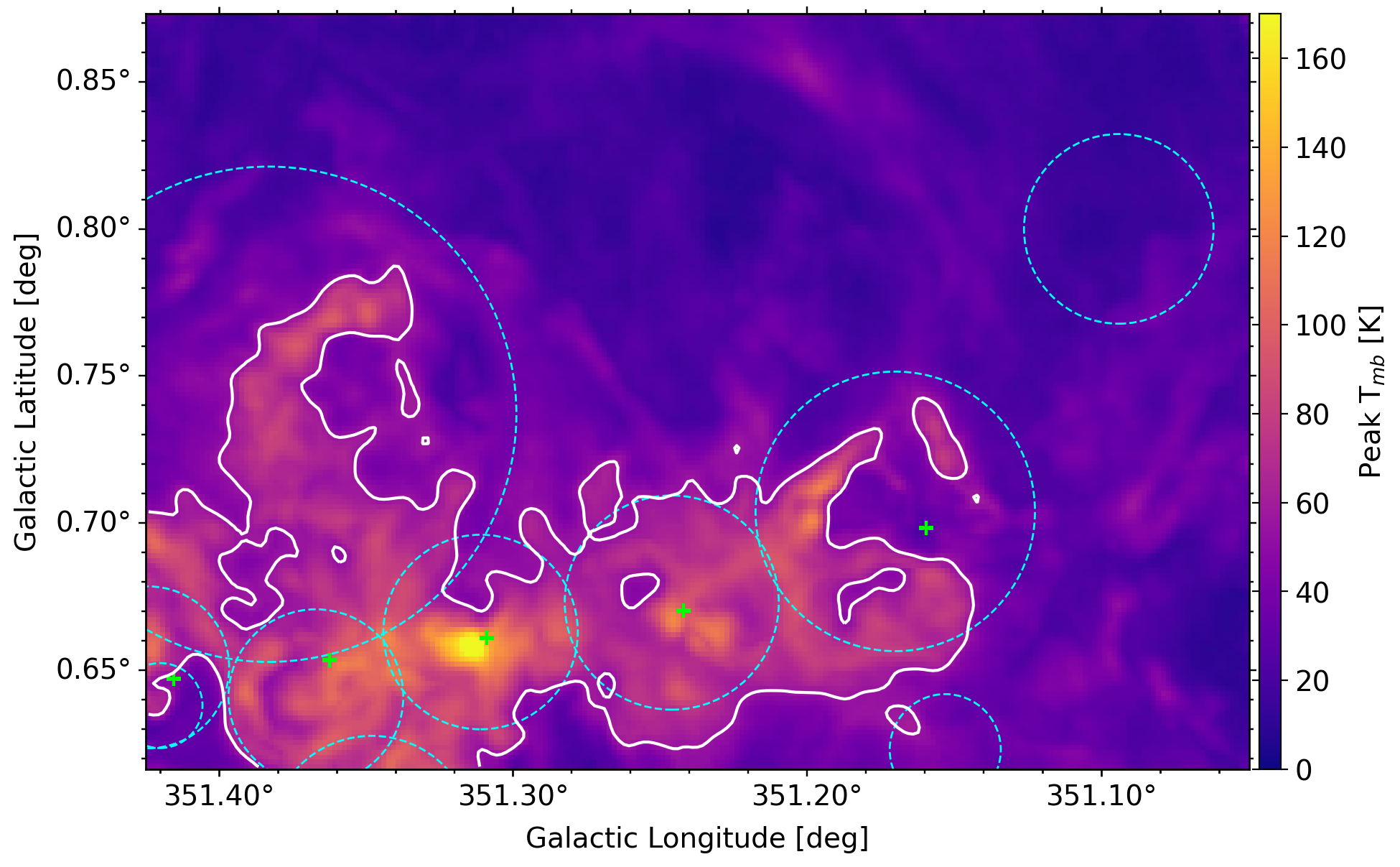}
    \caption{\CII peak emission map. The contour is drawn at T$_\mathrm{mb}$ of 55 K.}
    \label{fig:ciitmbpeak}
\end{figure*}

\begin{figure*}[htbp!]
    \centering
    \includegraphics[width=0.45\linewidth]{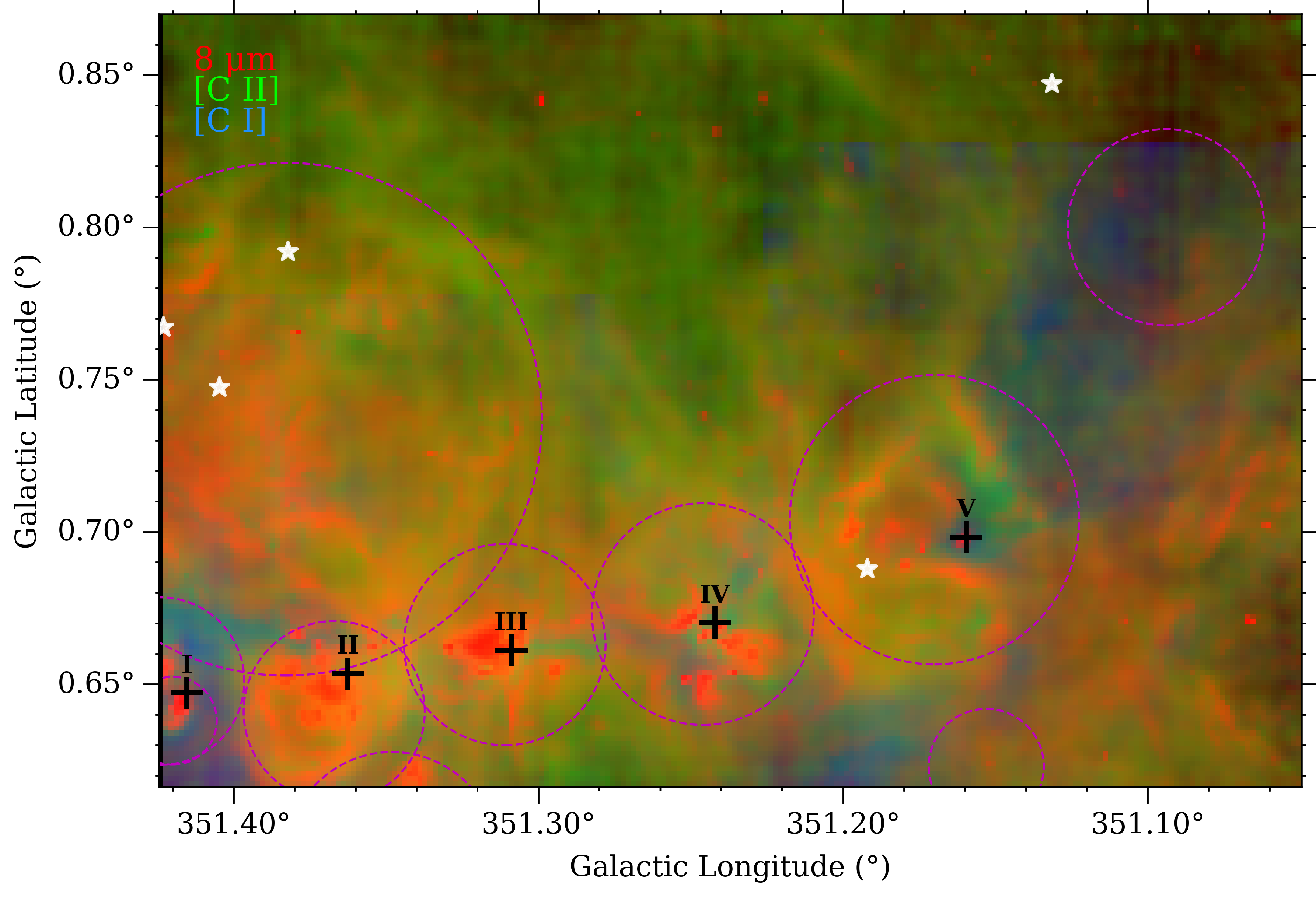}
    \includegraphics[width=0.45\linewidth]{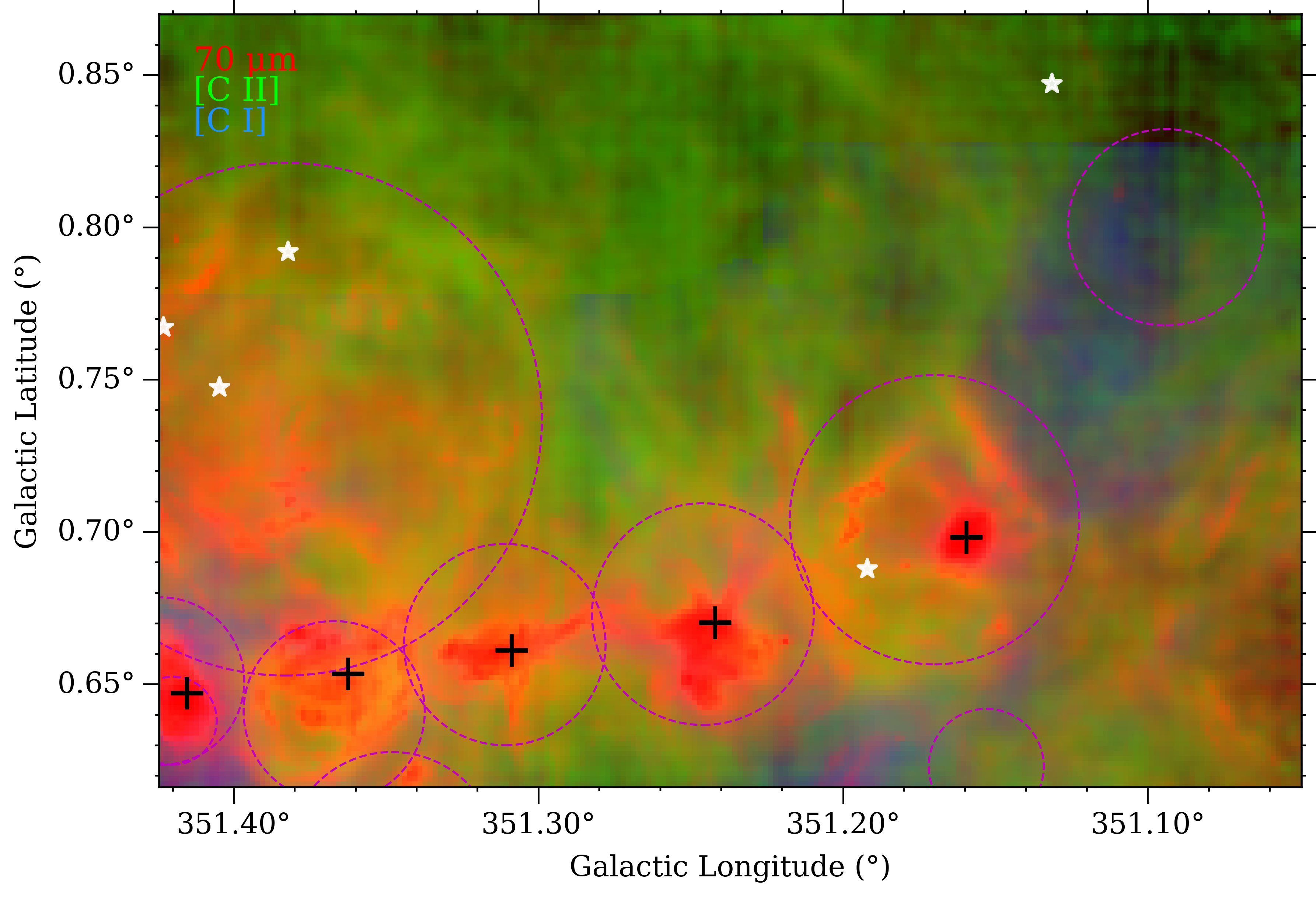}
    \caption{Three color images of the NGC~6334 central filament. Left: 8 $\mu$m (red), \cCII (green), \cCI (blue). Right: 70 $\mu$m (red), \cCII (green), \cCI (blue). The symbols and markers are same as in Fig. \ref{fig:mom0map1}, except that the OB--type stars are shown in white color.} 
    \label{fig:3color_with_8_70mic}
\end{figure*}

\begin{figure}[htbp!]
\sidecaption
    \includegraphics[width=12cm]{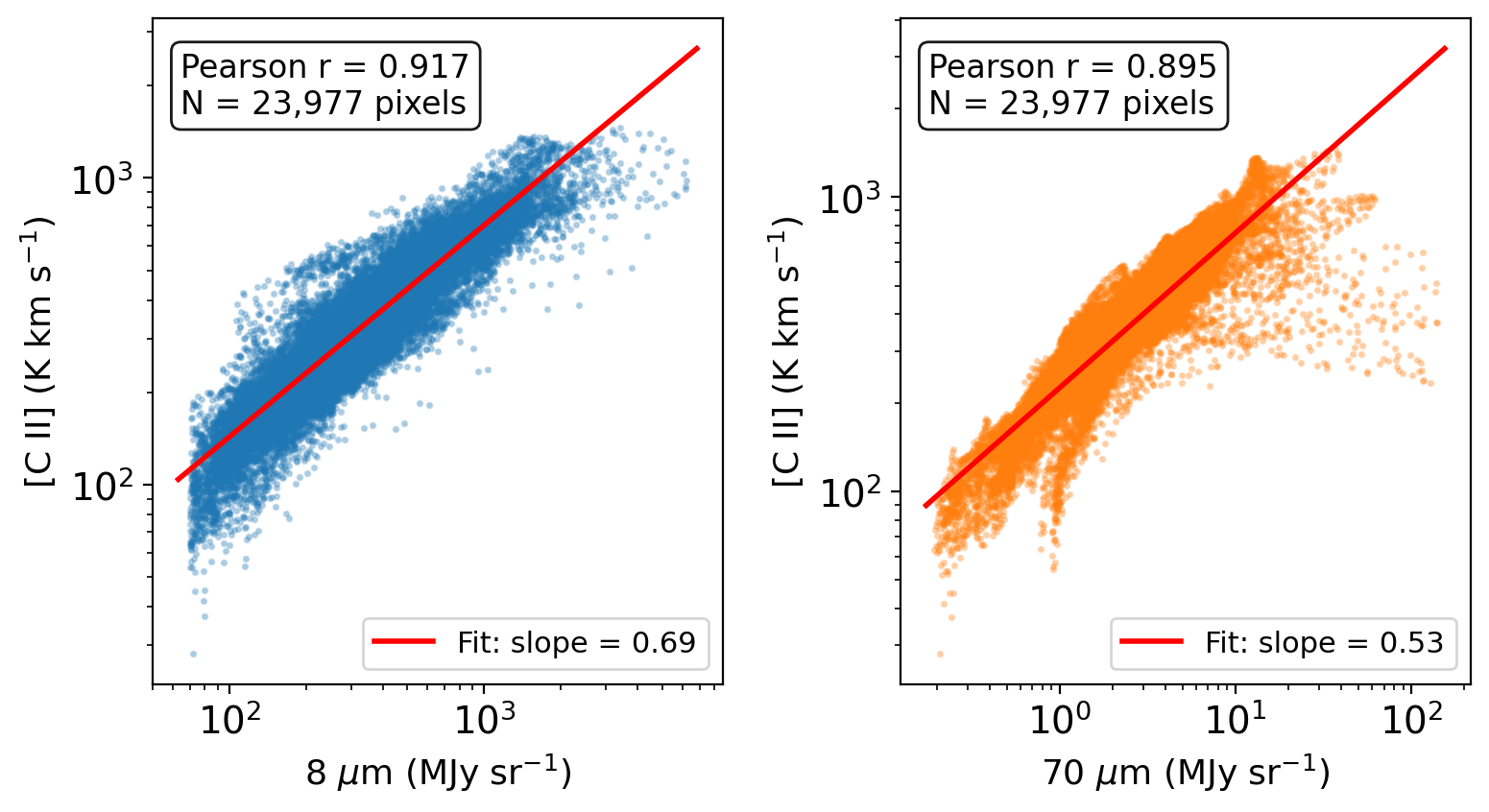}
    \caption{\cCII versus dust correlations (left: with 8 $\mu$m, right: with 70 $\mu$m).}
    \label{fig:cii_vs_dust}
\end{figure}

\onecolumn
\begin{figure*}[htbp!]
    \centering
        \includegraphics[width=0.9\linewidth]{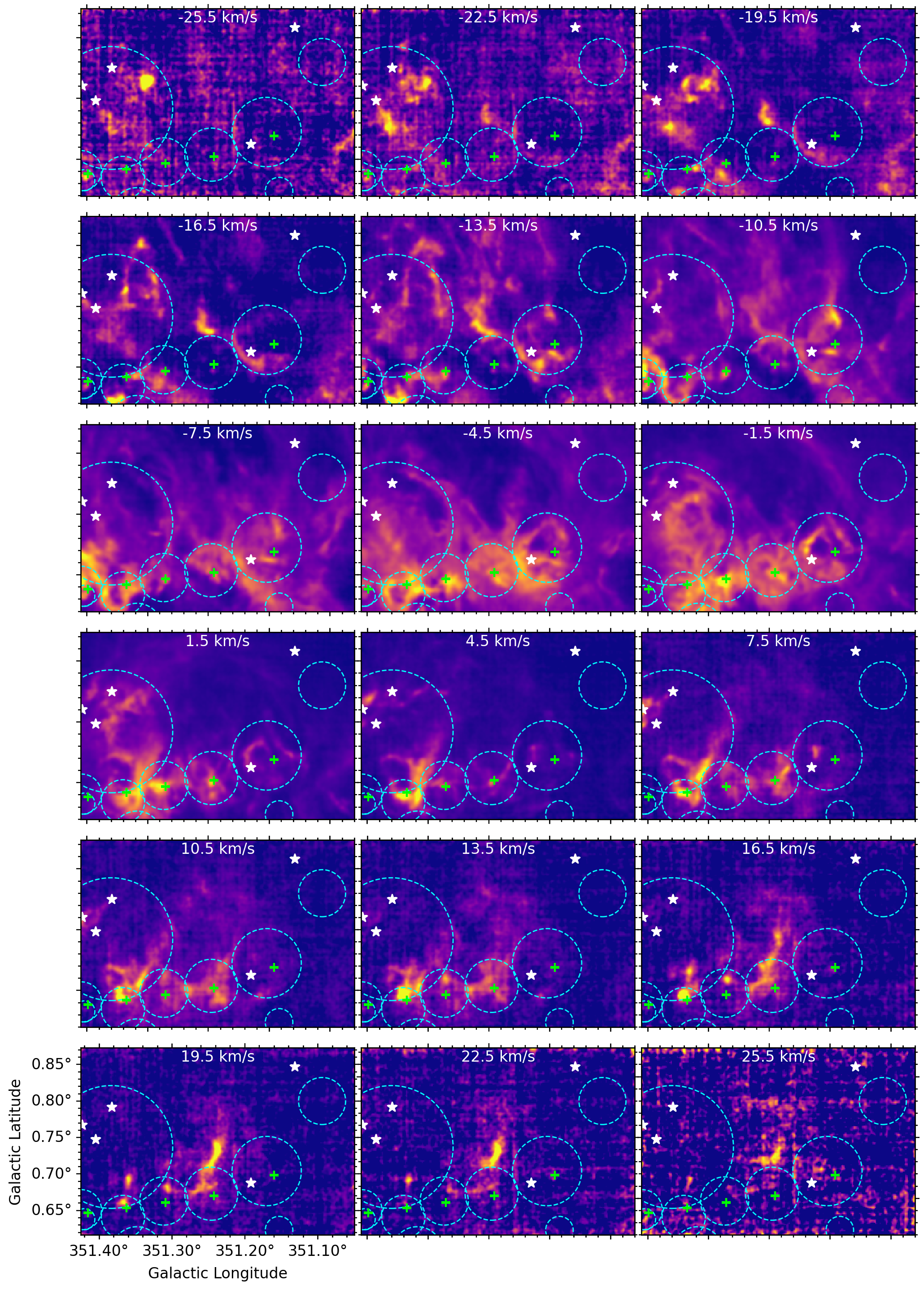}
    \caption{Channel maps of \CII emission from -27 to +27 \kms with a step of 3 \kms. 
    Other symbols are same as in Fig. \ref{momentmaps}.}
    \label{fig:cii_channelmap}
\end{figure*}

\begin{figure*}[htbp!]
    \centering
    \includegraphics[width=0.75\linewidth]{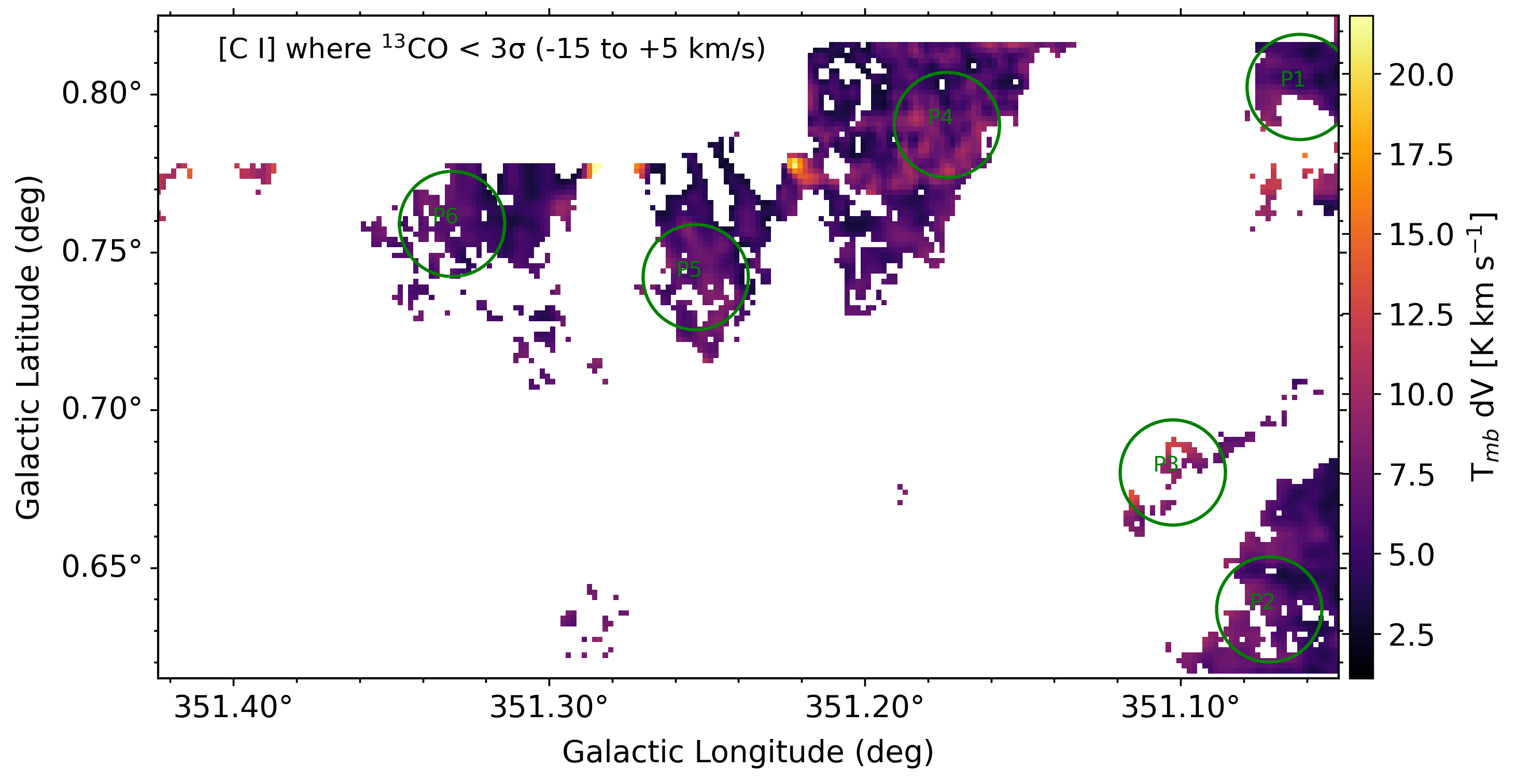}\\
    \includegraphics[width=0.75\linewidth]{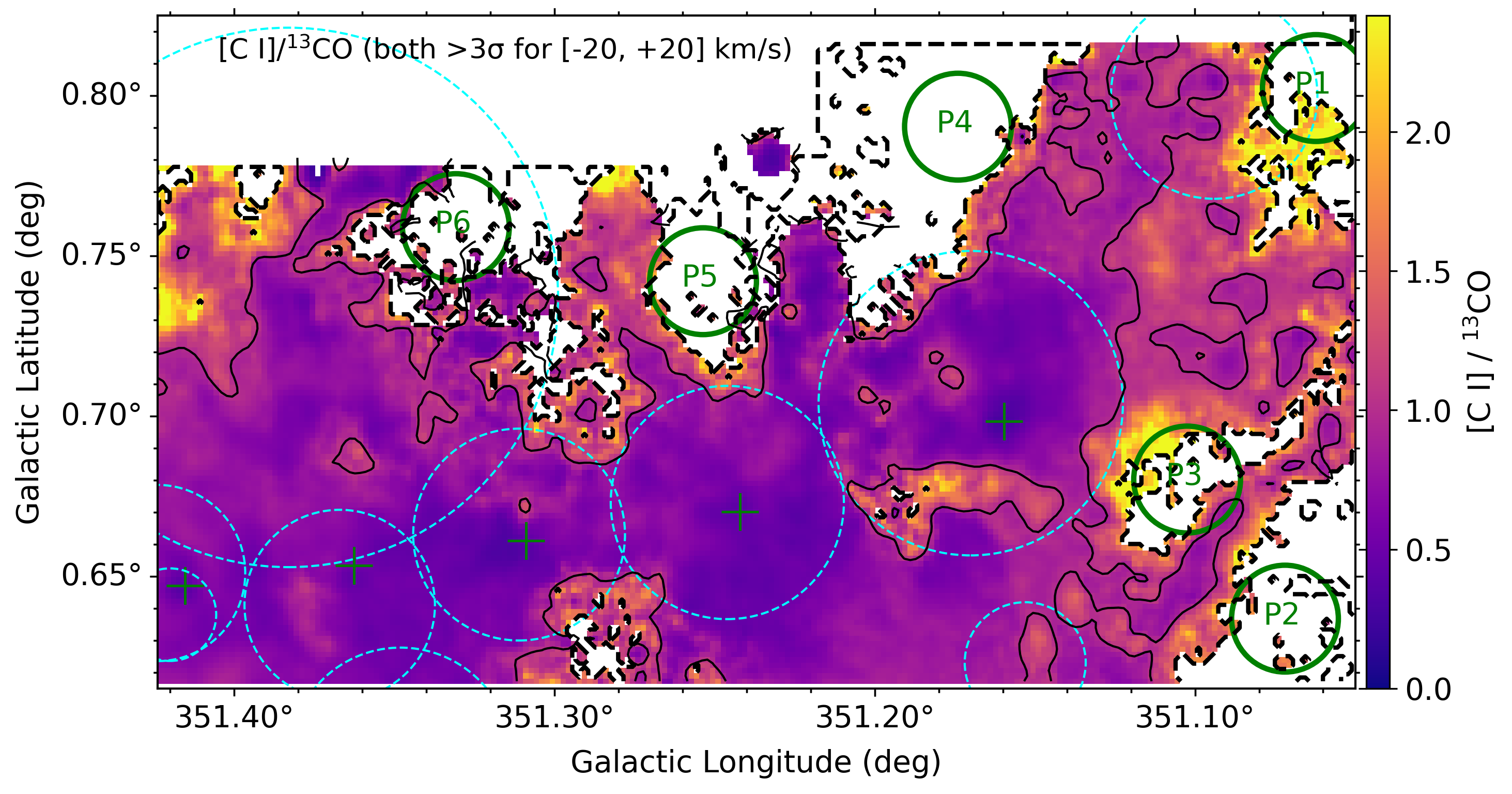}\\
    \includegraphics[width=0.7\linewidth]{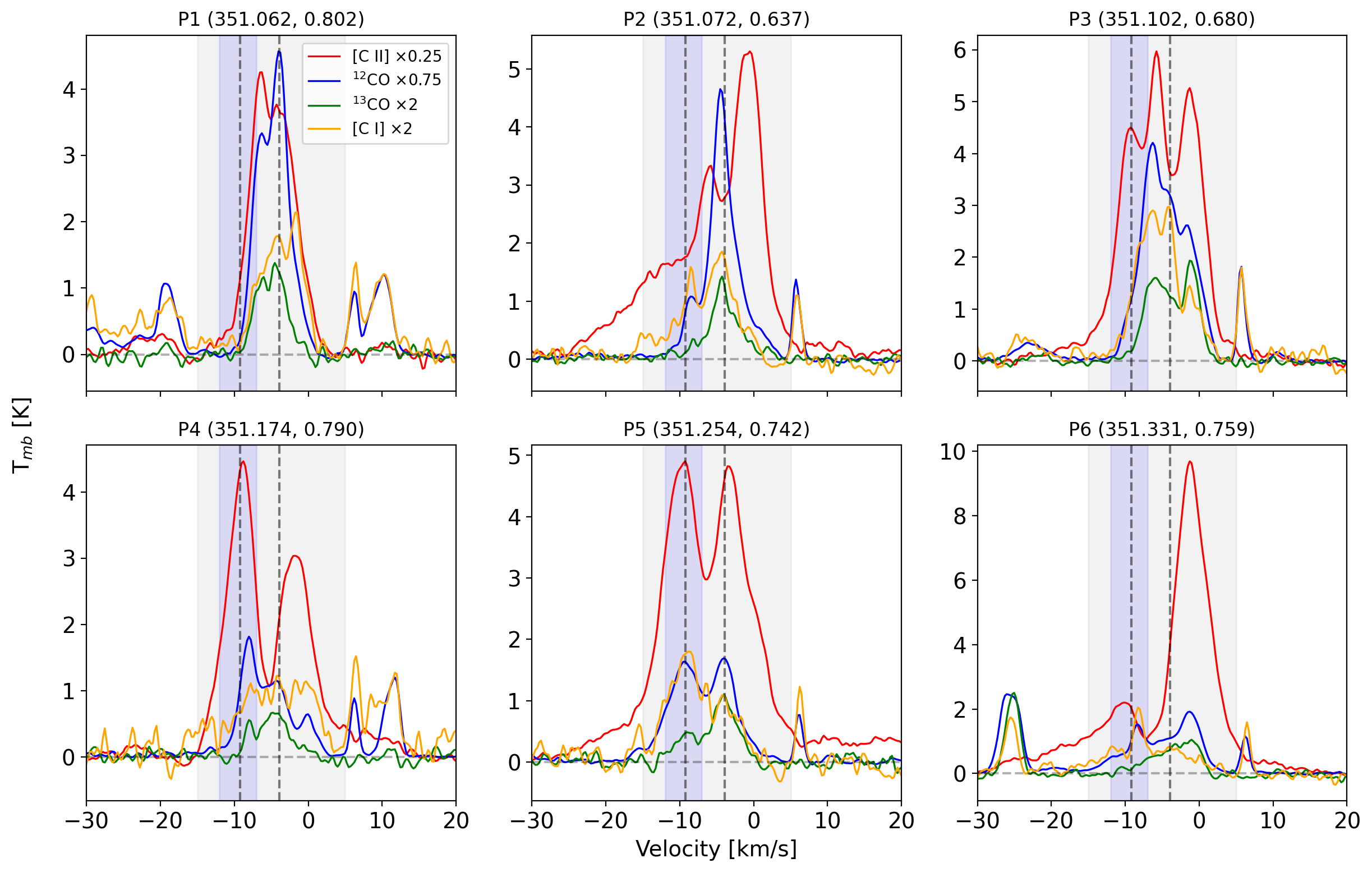}
    \caption{Top: Map of \CI integrated intensity (K km s$^{-1}$) for pixels where CI $>$ 3$\sigma$ and $^{13}$CO $<$ 3$\sigma$, in velocity ranges -15 to +5 \kms. Middle: CI/$^{13}$CO ratio map in velocity ranges -20 to +20 \kms . Contour level at the ratio 1 is shown in black color and mask region where CI $>$ 3$\sigma$ and $^{13}$CO $<$ 3$\sigma$ (created from the integrated intensity maps) is shown by the black dotted contour lines. Bottom: Average spectra toward regions P1 to P6 indicated by green circles (of size 60$^{''}$) in the top and middle panel. Black dotted lines indicate two main velocity components (-3.9 and -9.2 \kms) of $^{13}$CO toward the region \citep{Neupane2024}}
    \label{fig:ci_not_13co}
\end{figure*}

\onecolumn
\section{Position velocity maps of \CII emission associated with \HII regions}
\FloatBarrier
\begin{figure*}[htbp!]
    \centering
    \includegraphics[width=0.49\linewidth, trim = 2cm 1cm 2cm 3cm, clip]{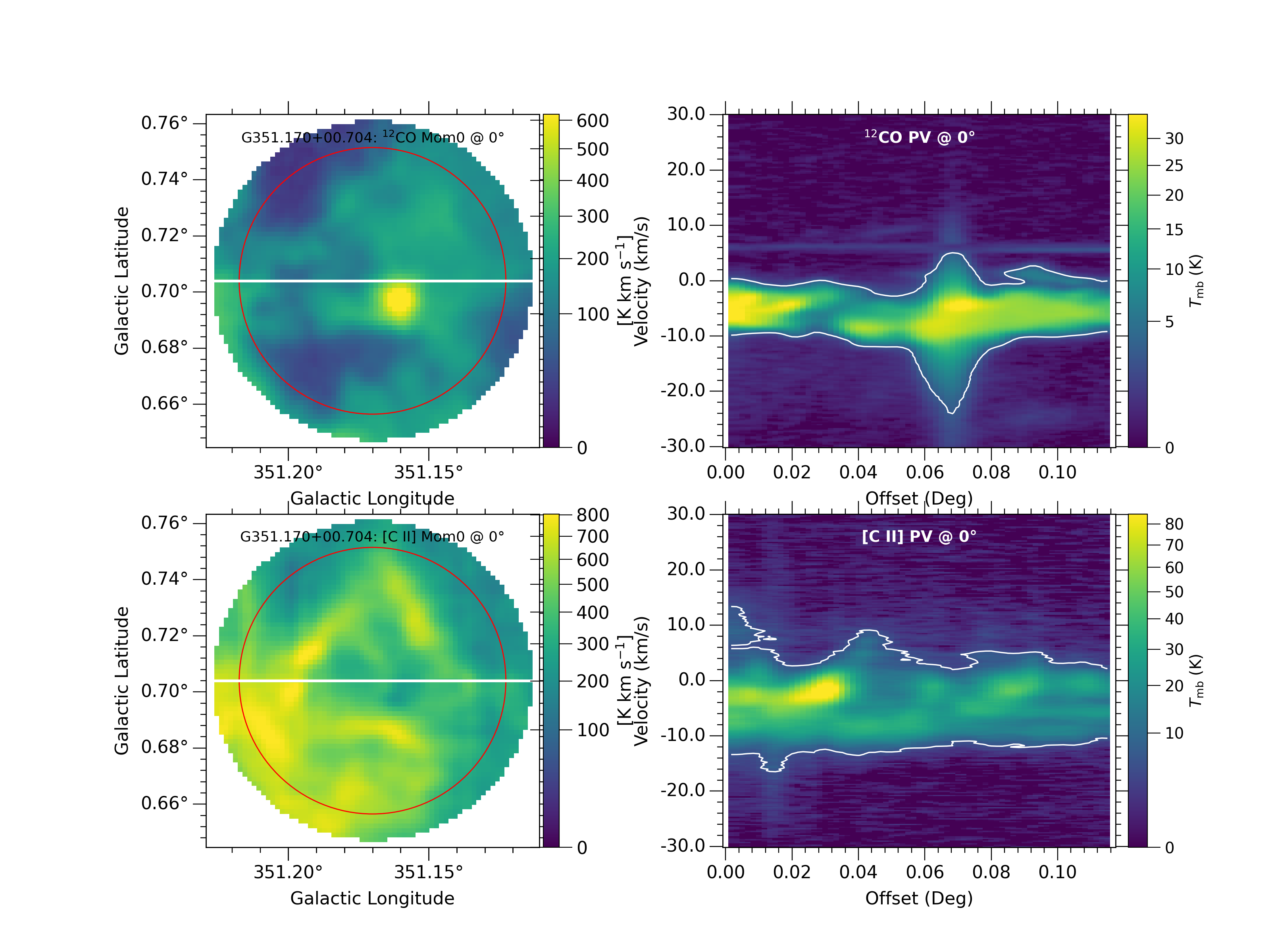}
    \includegraphics[width=0.49\linewidth, trim = 2cm 1cm 2cm 3cm, clip]{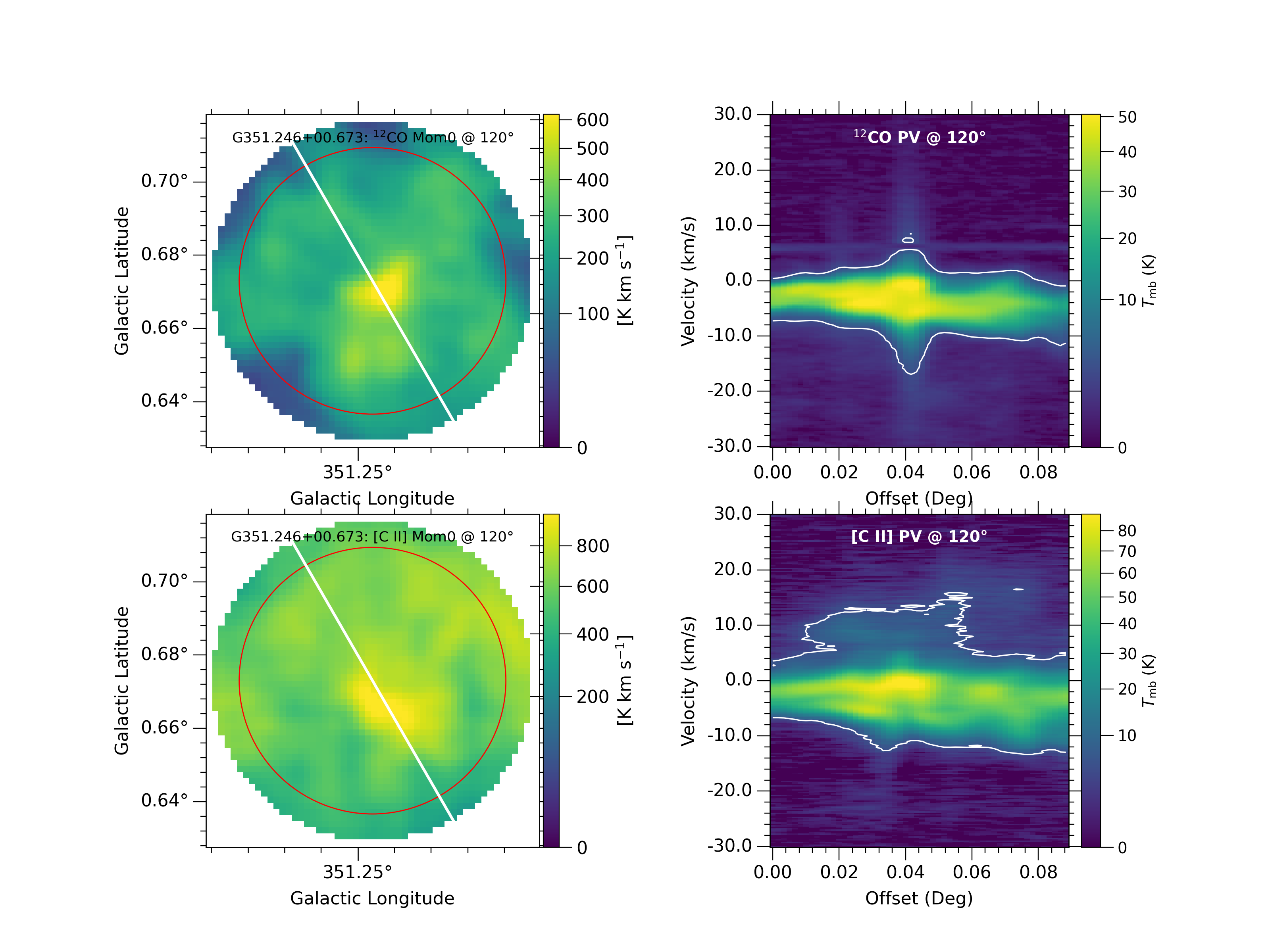}\\
    \includegraphics[width=0.49\linewidth, trim = 2cm 1cm 2cm 3cm, clip]{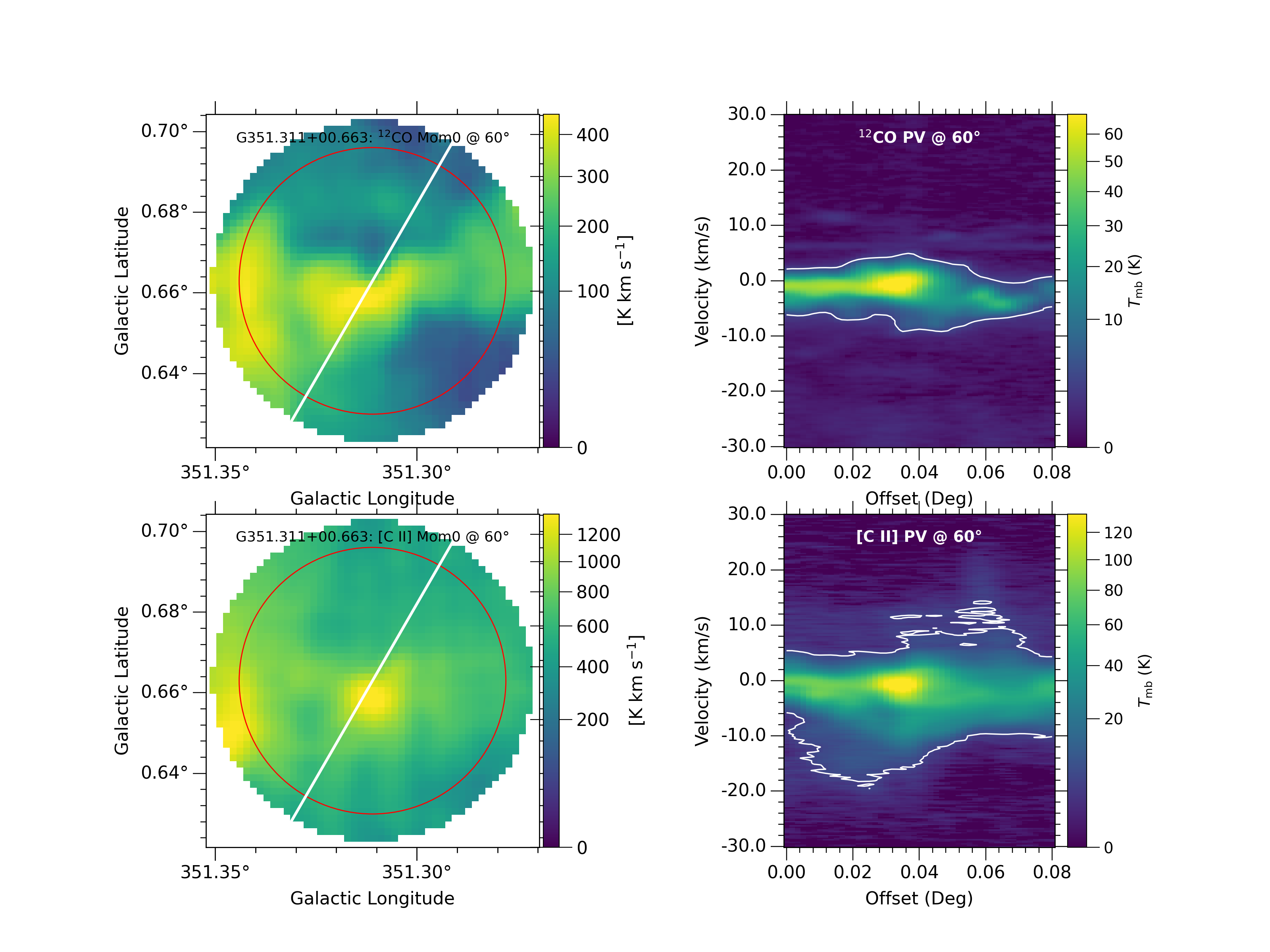}
    \includegraphics[width=0.49\linewidth, trim = 2cm 1cm 2cm 3cm, clip]{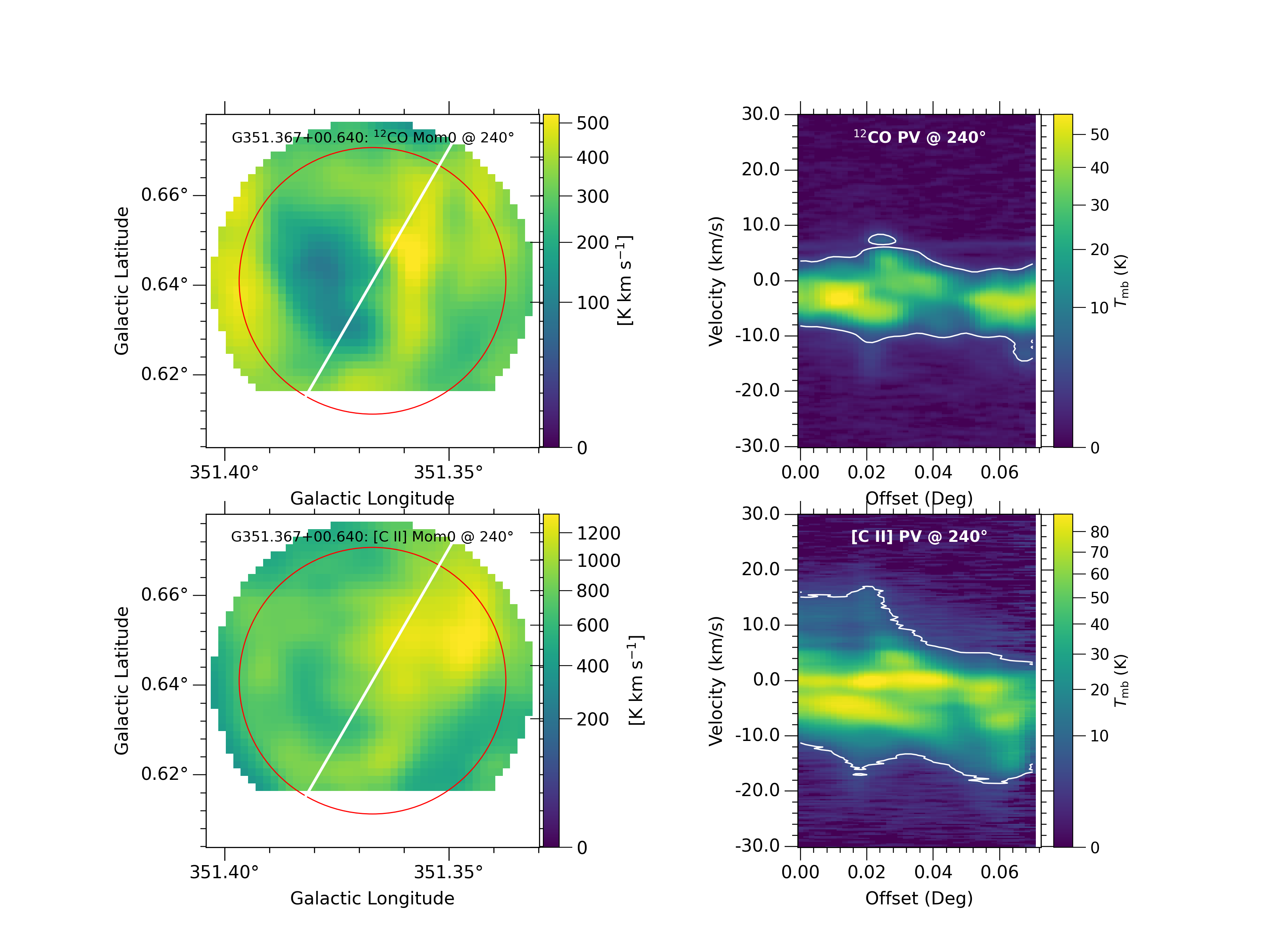}\\
    \includegraphics[width=0.49\linewidth, trim = 2cm 1cm 2cm 3cm, clip]{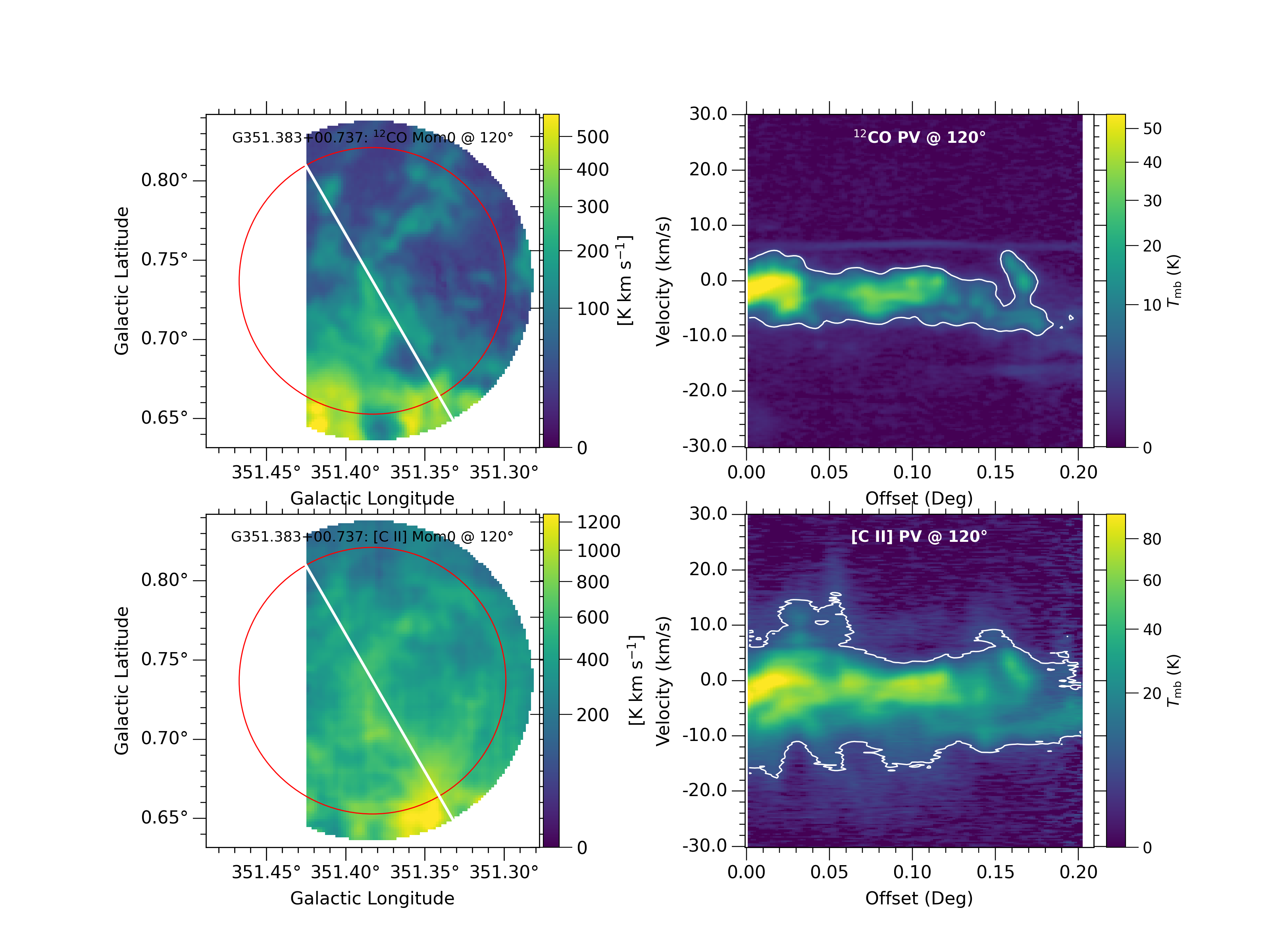}
    \caption{Position-velocity maps along different angles passing through the center of the \HII regions. On each figure, left panels show moment-0 maps of $^{12}$CO and \CII. On the right panels, PV-maps are constructed along the path shown in the moment-0 maps. Contours are drawn at 5$\sigma$ noise level on each maps. Movies of the PV-maps for rotating angular slices are available in the online versions.} \label{pvmaps_different_angles}
\end{figure*}
 
\FloatBarrier
\section{Figures and tables}
\begin{figure*}[htbp!]
    \centering
    \includegraphics[width=0.4\linewidth]{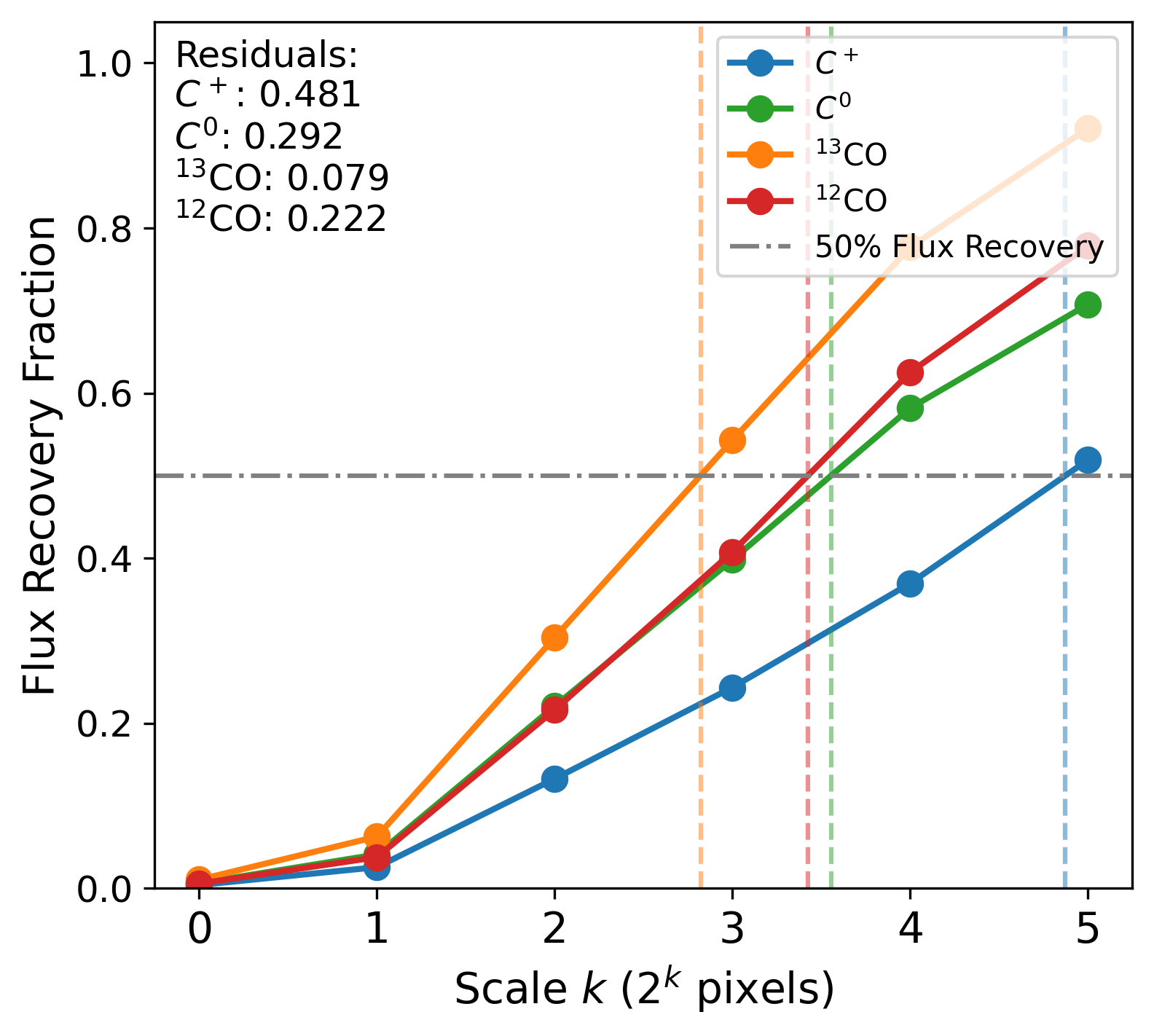}      
    \includegraphics[width=0.4\linewidth]{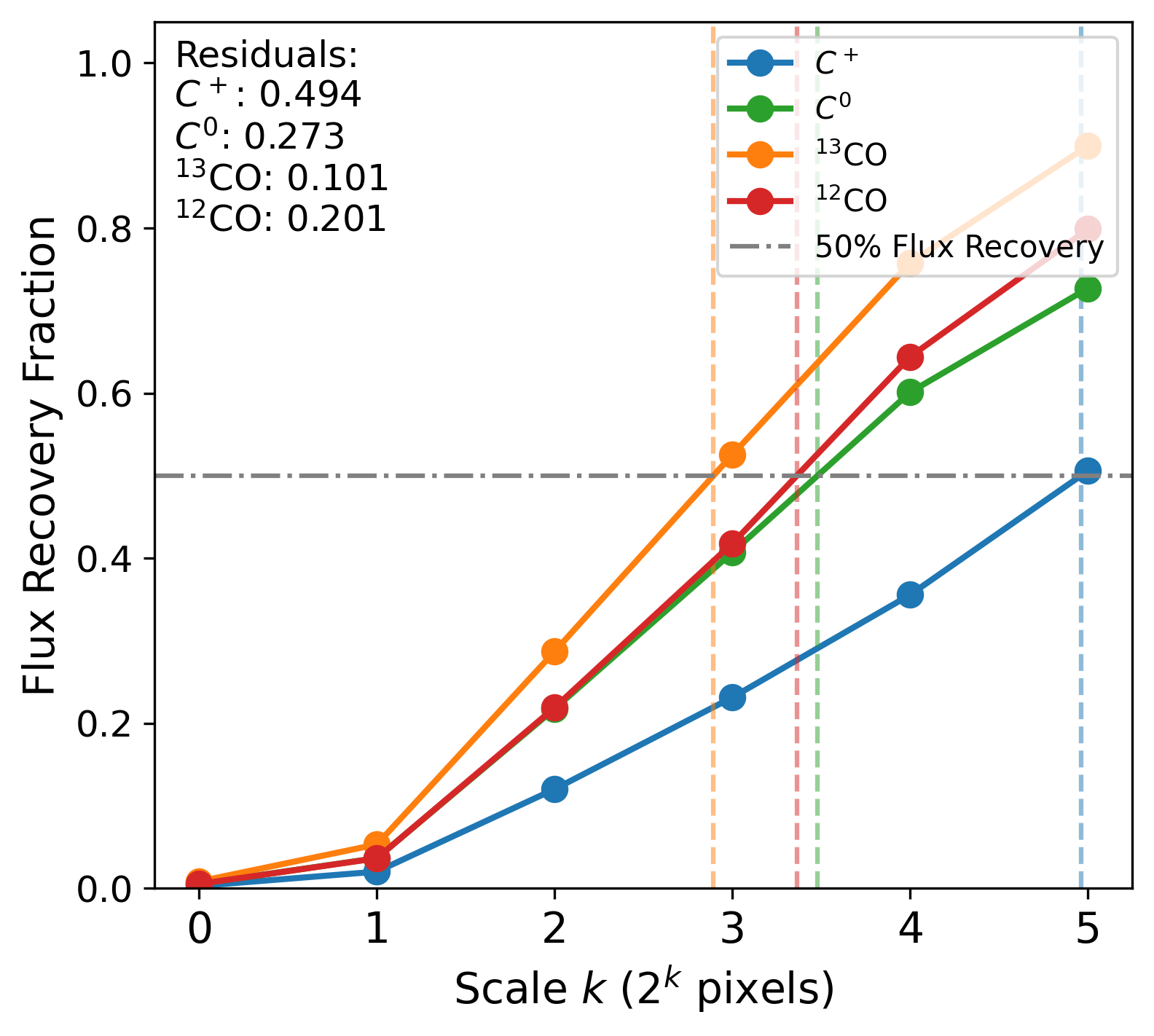}

    \caption{Flux recovery at different scales obtained from the constrained diffusion decomposition (CDD) model. The  moment-0 maps were used for the velocity ranges -30 to +30 \kms (left) and -15 to +5 \kms (right). Residual fractions are also noted for each species.}
    \label{fig:cdd}
\end{figure*}

\begin{figure}[htbp!]
    \centering
    \begin{subfigure}{0.45\textwidth}
        \centering
        \includegraphics[width=0.8\linewidth]{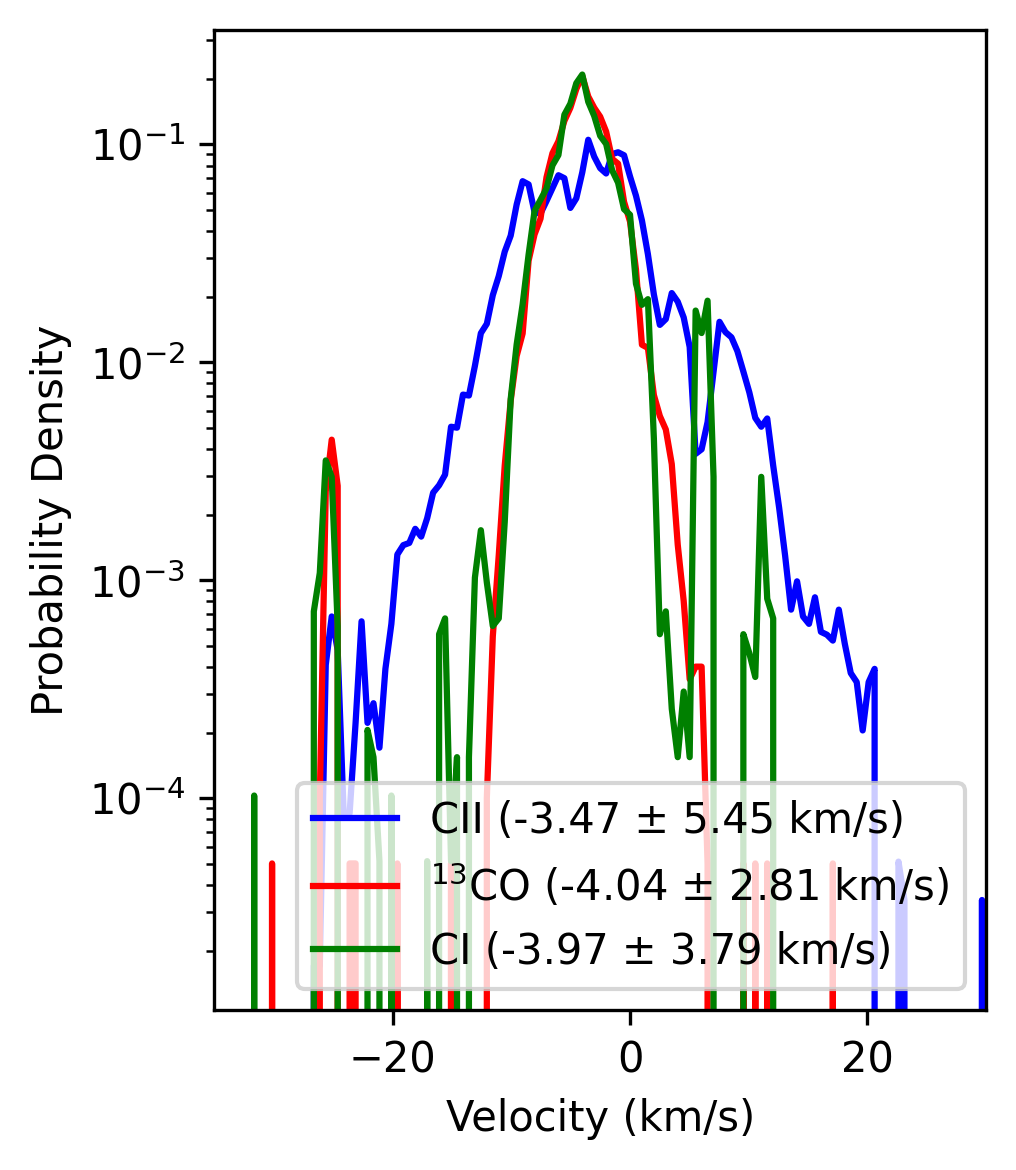}
        \caption{}
        \label{fig:sub_a}
    \end{subfigure}
    \begin{subfigure}{0.45\textwidth}
        \centering
        \includegraphics[width=0.85\linewidth]{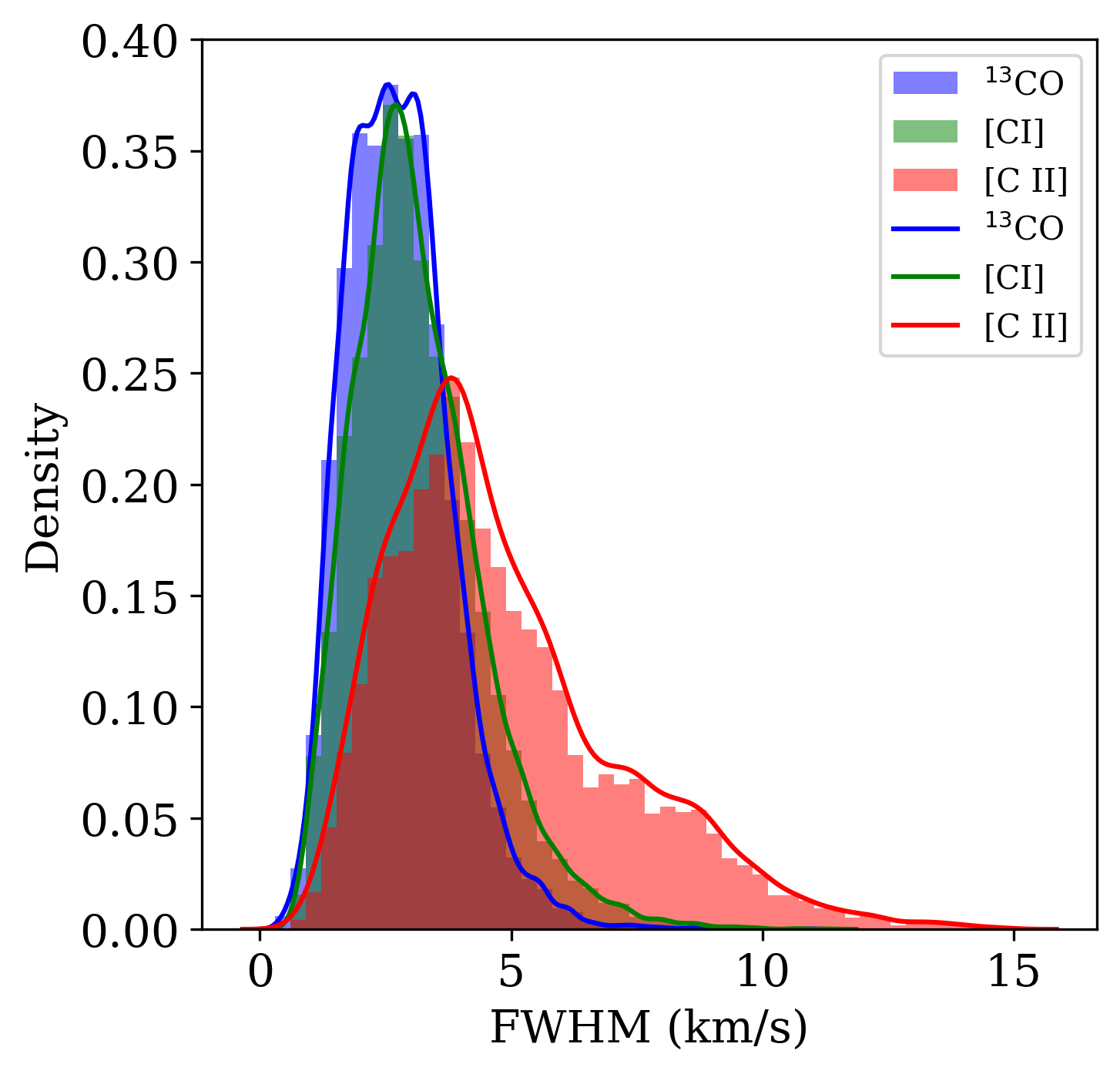}
        \caption{}
        \label{fig:sub_b}
    \end{subfigure}
    \caption{(a) Velocity centroid PDF made using the Gauss fit components. (b) Line widths distribution of the three species from Gauss fit peak component analysis. The color lines over plotted on the histograms are kernel density estimation (KDE) lines. }
    \label{fig:velpdf_and_lwdist}
\end{figure}

\begin{figure}[htbp!]
    \sidecaption
    \includegraphics[width=12cm]{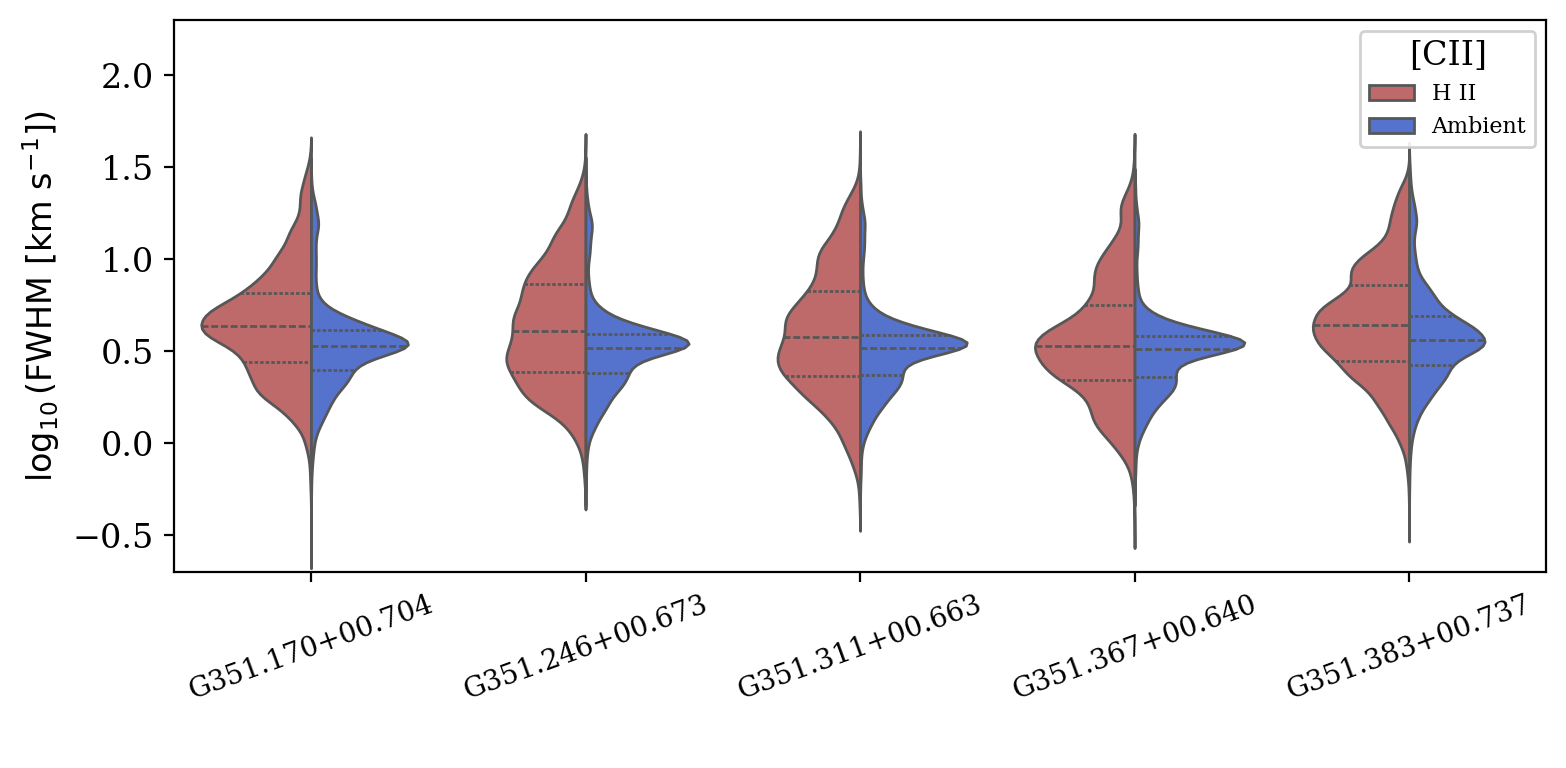}
    \caption{Linewidth comparison of \HII regions (in red) with respect to ambient region (in blue). The ambient region is centered at $l=351.081^{\circ}$ and $b=0.745^{\circ}$ westward to the FIR-V, with dynamical size matching the size of each \HII regions. Though average values of the linewidth are similar to ambient region, all \HII regions exhibit high linewidth tail.}
    \label{fig:lw_hii_vs_ambient}
\end{figure}

\begin{figure}[htbp!]
    \centering
    \sidecaption
    \includegraphics[width=8cm]{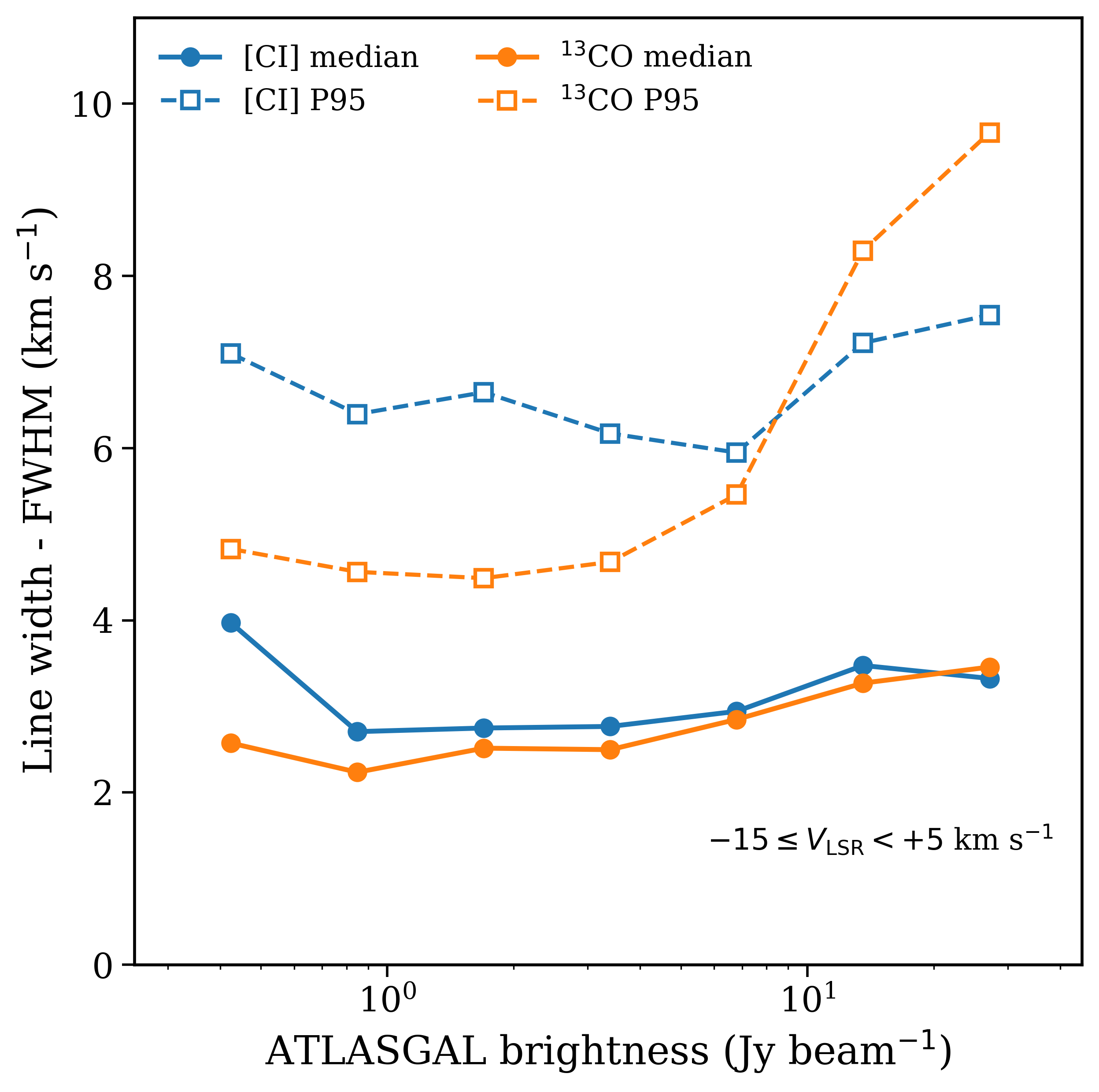}
    \caption{\CI and $^{13}$CO linewidth in logarithmic ATLASGAL 870 $\mu$m brightness bins. P95 indicate 95th percentile linewidth for each bins.} \label{fig:ci_13co_linewidth}
\end{figure}

\begin{figure*}[htb!]
    \sidecaption
    \includegraphics[width=12cm]{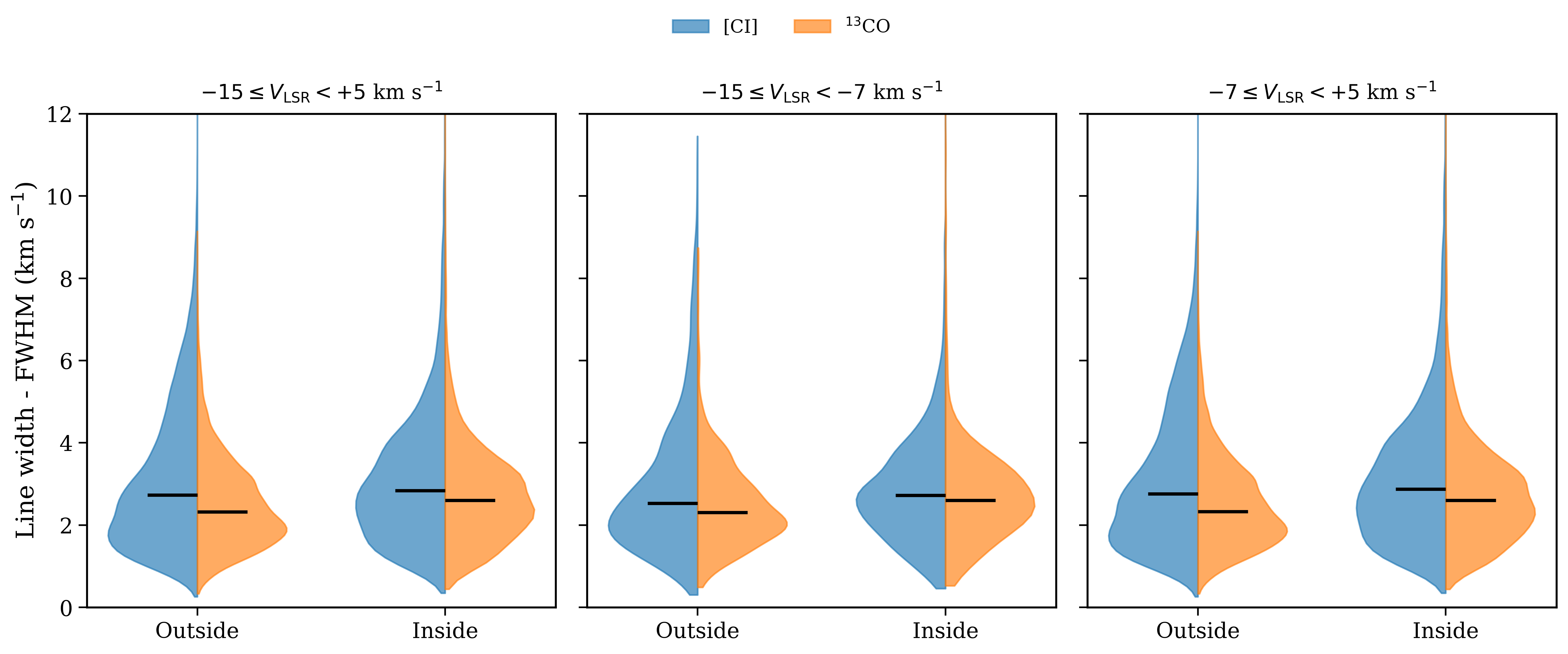}
    \caption{\CI and $^{13}$CO linewidth distribution for Gauss components in different velocity ranges toward Outside vs Inside region. } \label{fig:ci_13co_linewidth_violin}
\end{figure*}

\begin{figure*}[htbp!]
\sidecaption
    \includegraphics[width=12cm]{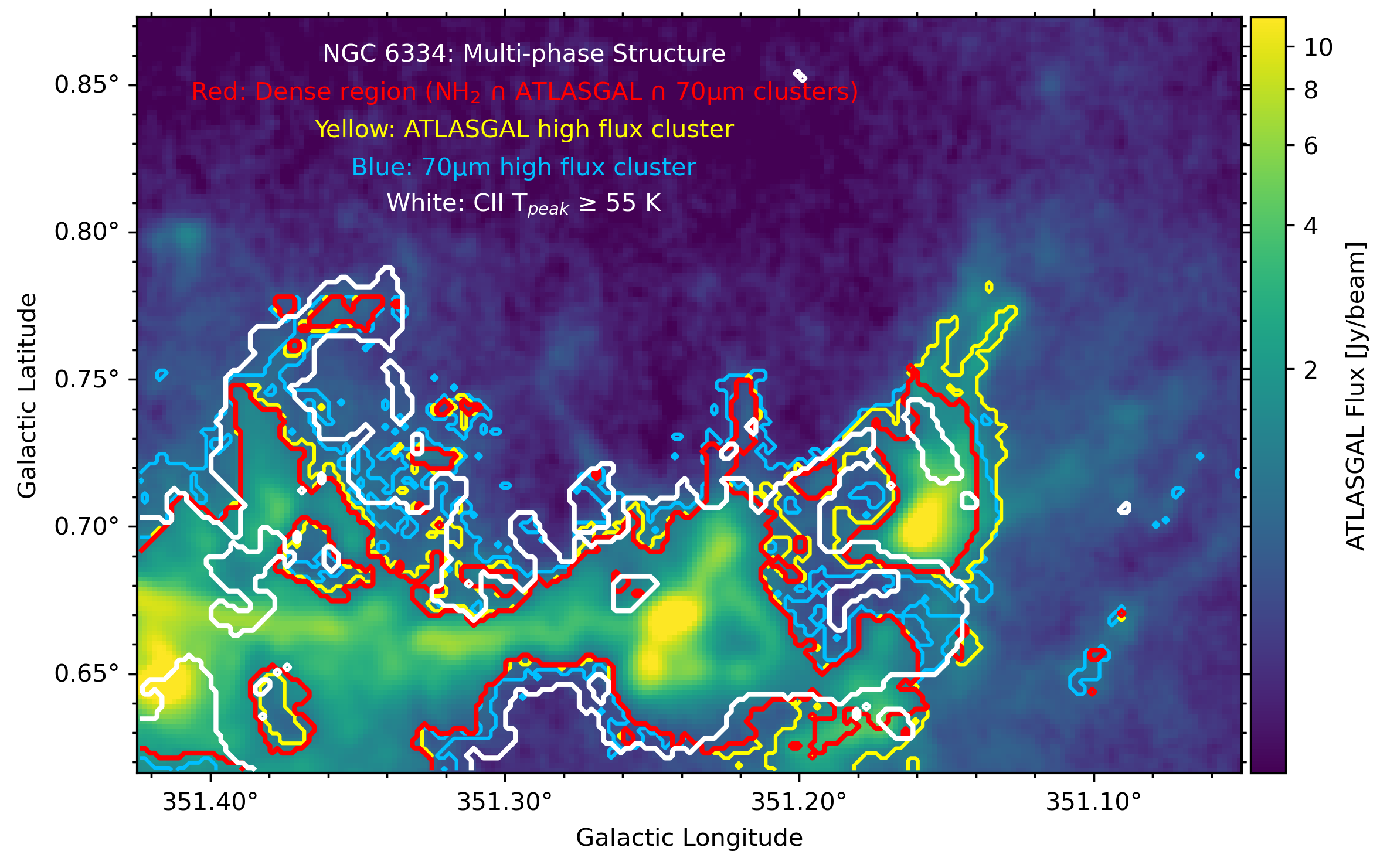}
    \caption{ATLASGAL map overlayed with  \cCII intensity contour at 55\,K showing the bright [C II] region. High flux cluster boundaries based on N(H$_{2}$), ATLASGAL, and 70 $\mu$m emission are shown in red, yellow and blue color, respectively. }
    \label{fig:clustermasks}
\end{figure*}

\FloatBarrier
\onecolumn
\begin{table*}[htbp!]
\footnotesize
\centering
\caption{Statistics of \CII, $^{13}$CO, and \CI Gaussian components.}
\label{tab:general_stats_threepanels}
\begin{tabular}{l|l|cccccc}
\hline\hline
\text{Tracer} & \text{Parameter} & \text{Mean} & \text{Median} & \text{Std. Dev.} & \text{Skewness} & \text{Kurtosis} \\
\hline
\multicolumn{7}{c}{\text{Full Map}} \\
\hline
\multirow{3}{*}{\CII} 
  & V (\kms) & -3.47 & -3.45 & 5.45 & 0.40 & 1.19 \\
  & T$_{\rm mb}$ (K) & 18.10 & 13.21 & 15.55 & 1.84 & 4.23 \\
  & $\Delta V$ (\kms) & 5.42 & 4.22 & 4.22 & 2.45 & 8.02 \\
\hline
\multirow{3}{*}{$^{13}$CO} 
  & V (\kms) & -4.03 & -4.01 & 2.80 & -1.97 & 15.11 \\
  & T$_{\rm mb}$ (K) & 8.71 & 6.96 & 6.41 & 1.39 & 2.10 \\
  & $\Delta V$ (\kms) & 2.60 & 2.42 & 1.19 & 1.64 & 9.33 \\
\hline
\multirow{3}{*}{\CI} 
  & V (\kms) & -3.94 & -4.09 & 3.61 & -1.57 & 19.65 \\
  & T$_{\rm mb}$ (K) & 5.17 & 3.80 & 4.06 & 1.68 & 3.14 \\
  & $\Delta V$ (\kms) & 3.06 & 2.73 & 1.73 & 1.20 & 2.32 \\
\hline
\multicolumn{7}{c}{\text{Inside Region (\CII $T_{\rm peak} \geq 55$\,K)}} \\
\hline
\multirow{3}{*}{\CII} 
  & V (\kms) & -1.92 & -2.41 & 6.07 & 0.21 & 0.56 \\
  & T$_{\rm mb}$ (K) & 24.01 & 16.66 & 20.78 & 1.14 & 0.93 \\
  & $\Delta V$ (\kms) & 5.51 & 3.88 & 4.79 & 2.15 & 5.42 \\
\hline
\multirow{3}{*}{$^{13}$CO} 
  & V (\kms) & -3.26 & -3.17 & 2.61 & -1.08 & 8.80 \\
  & T$_{\rm mb}$ (K) & 10.80 & 8.49 & 7.99 & 0.99 & 0.39 \\
  & $\Delta V$ (\kms) & 2.69 & 2.50 & 1.18 & 1.28 & 4.45 \\
\hline
\multirow{3}{*}{\CI} 
  & V (\kms) & -3.44 & -3.40 & 3.40 & -3.27 & 34.41 \\
  & T$_{\rm mb}$ (K) & 5.78 & 4.38 & 4.35 & 1.36 & 1.65 \\
  & $\Delta V$ (\kms) & 3.20 & 2.81 & 1.84 & 1.42 & 3.13 \\
\hline
\multicolumn{7}{c}{\text{Outside Region (\CII $T_{\rm peak} < 55$\,K)}} \\
\hline
\multirow{3}{*}{\CII} 
  & V (\kms) & -3.98 & -3.94 & 5.06 & 0.58 & 2.22 \\
  & T$_{\rm mb}$ (K) & 16.17 & 14.18 & 10.51 & 0.87 & 0.22 \\
  & $\Delta V$ (\kms) & 5.50 & 4.11 & 4.42 & 2.47 & 7.72 \\
\hline
\multirow{3}{*}{$^{13}$CO} 
  & V (\kms) & -4.73 & -4.56 & 2.73 & -2.64 & 19.04 \\
  & T$_{\rm mb}$ (K) & 7.24 & 6.15 & 4.71 & 1.07 & 0.91 \\
  & $\Delta V$ (\kms) & 2.56 & 2.38 & 1.26 & 2.47 & 17.77 \\
\hline
\multirow{3}{*}{\CI} 
  & V (\kms) & -4.33 & -4.58 & 4.55 & -1.49 & 14.51 \\
  & T$_{\rm mb}$ (K) & 4.05 & 2.93 & 3.01 & 1.78 & 3.99 \\
  & $\Delta V$ (\kms) & 2.91 & 2.57 & 1.69 & 1.03 & 1.17 \\
\hline
\end{tabular}
\end{table*}

\begin{table}[htbp!]
\footnotesize
\centering
\caption{Statistics of line ratios derived from moment-0 maps.}
\label{tab:line_ratios}
\begin{tabular}{l|c|c|c|c|c}
\hline\hline
\multicolumn{6}{c}{Full Map} \\\hline
Ratio & Mean & Median & Std. Dev. & Skewness & Kurtosis \\
\hline
$\mathrm{[CII]}/^{12}\mathrm{CO}$ & 4.79 & 3.28 & 3.86 & 1.69 & 3.37 \\
$\mathrm{[CII]}/^{13}\mathrm{CO}$ & 18.25 & 12.34 & 15.95 & 1.61 & 2.85 \\
$\mathrm{[CII]}/\mathrm{[CI]}$    & 20.05 & 16.47 & 13.77 & 1.28 & 1.68 \\
$\mathrm{[CI]}/^{13}\mathrm{CO}$  & 0.97 & 0.85 & 0.46 & 1.95 & 7.02 \\
$\mathrm{[CI]}/^{12}\mathrm{CO}$  & 0.24 & 0.22 & 0.10 & 6.43 & 105.63 \\
$^{12}\mathrm{CO}/^{13}\mathrm{CO}$ & 4.57 & 4.22 & 1.73 & 1.10 & 1.42 \\
\hline
\multicolumn{6}{c}{Inside $T_{\rm peak}[\mathrm{CII}] \geq 55$\,K} \\
\hline
Ratio & Mean & Median & Std & Skewness & Kurtosis \\
\hline
$\mathrm{[CII]}/^{12}\mathrm{CO}$ & 3.97 & 3.00 & 2.76 & 2.26 & 6.54 \\
$\mathrm{[CII]}/^{13}\mathrm{CO}$ & 16.78 & 10.71 & 15.51 & 2.17 & 5.93 \\
$\mathrm{[CII]}/\mathrm{[CI]}$ & 20.51 & 16.71 & 12.83 & 1.28 & 1.48 \\
$\mathrm{[CI]}/^{13}\mathrm{CO}$ & 0.76 & 0.70 & 0.28 & 1.89 & 5.35 \\
$\mathrm{[CI]}/^{12}\mathrm{CO}$ & 0.20 & 0.19 & 0.05 & 0.69 & 0.24 \\
$^{12}\mathrm{CO}/^{13}\mathrm{CO}$ & 3.88 & 3.57 & 1.27 & 1.16 & 1.62 \\
\hline
\multicolumn{6}{c}{Outside $T_{\rm peak}[\mathrm{CII}] < 55$\,K} \\
\hline
Ratio & Mean & Median & Std & Skewness & Kurtosis \\
\hline
$\mathrm{[CII]}/^{12}\mathrm{CO}$ & 5.28 & 3.69 & 4.31 & 1.38 & 2.08 \\
$\mathrm{[CII]}/^{13}\mathrm{CO}$ & 19.45 & 14.54 & 16.20 & 1.21 & 1.00 \\
$\mathrm{[CII]}/\mathrm{[CI]}$ & 19.76 & 16.26 & 14.33 & 1.28 & 1.73 \\
$\mathrm{[CI]}/^{13}\mathrm{CO}$ & 1.15 & 1.05 & 0.50 & 1.73 & 6.01 \\
$\mathrm{[CI]}/^{12}\mathrm{CO}$ & 0.26 & 0.23 & 0.12 & 6.16 & 88.35 \\
$^{12}\mathrm{CO}/^{13}\mathrm{CO}$ & 5.13 & 4.82 & 1.84 & 0.87 & 0.85 \\
\hline
\end{tabular}
\end{table}

\begin{table*}[htbp!]
\footnotesize
\centering
\caption{Line ratio statistics for the two K–means clusters.}
\label{tab:kmeans_stats}
\begin{tabular}{l|
                ccccc|c
                cccccc}
\hline
\multirow{2}{*}{\textbf{Ratio}} &
\multicolumn{5}{c|}{Cluster 1 (Low ATLASGAL flux)} &
\multicolumn{5}{c}{Cluster 2 (High ATLASGAL flux)}\\
\cline{2-6}\cline{7-11}
 & Mean & Median & Std & Skew & Kurtosis
 & Mean & Median & Std & Skew & Kurtosis \\
\hline
\(\mathrm{[C\,II]}/^{12}\mathrm{CO}\) & 4.45 & 3.43 & 3.24 & 1.84 & 4.59 & 2.23 & 2.19 & 0.90 & 1.00 & 2.96 \\
\(\mathrm{[C\,II]}/^{13}\mathrm{CO}\) & 23.33 & 18.93 & 16.71 & 1.27 & 1.83 & 6.96 & 6.27 & 3.67 & 1.68 & 5.06 \\
\(\mathrm{[C\,II]}/\mathrm{[C\,I]}\)  & 21.22 & 18.19 & 13.66 & 1.08 & 1.18 & 11.50 & 10.53 & 5.99 & 1.10 & 1.80 \\
\(\mathrm{[C\,I]}/^{13}\mathrm{CO}\) & 1.13 & 1.03 & 0.47 & 1.89 & 7.03 & 0.62 & 0.61 & 0.13 & 0.44 & 0.59 \\
\(\mathrm{[C\,I]}/^{12}\mathrm{CO}\) & 0.217 & 0.209 & 0.063 & 1.11 & 2.93 & 0.210 & 0.204 & 0.052 & 0.49 & -0.18 \\
\(^{12}\mathrm{CO}/^{13}\mathrm{CO}\) & 5.26 & 4.93 & 1.61 & 1.06 & 1.60 & 3.03 & 2.91 & 0.56 & 1.06 & 1.15 \\
\hline
\end{tabular}
\end{table*}

\begin{table}[htbp!]
\footnotesize
\centering
\caption{Line Intensity Ratio Statistics (Peak Component Analysis)}
\label{tab:ratio_stats_peakcomp_unclipped}
\begin{tabular}{l|ccccc}
\hline\hline
Ratio & Mean & Median & Std & Skew & Kurtosis \\
\hline
CI/$^{13}$CO & 0.63 & 0.52 & 0.51 & 5.51 & 76.84 \\
CII/CI & 17.64 & 10.47 & 25.06 & 5.32 & 52.98 \\
CII/$^{13}$CO & 9.68 & 5.12 & 13.61 & 4.40 & 37.24 \\
\hline
\end{tabular}
\end{table}

\end{appendix}

\end{document}